# The Design of the User Interfaces for Privacy Enhancements for Android

Jason I. Hong, Yuvraj Agarwal, Matt Fredrikson, Mike Czapik, Shawn Hanna, Swarup Sahoo, Judy Chun, Won-Woo Chung, Aniruddh Iyer, Ally Liu, Shen Lu, Rituparna Roychoudhury, Qian Wang, Shan Wang, Siqi Wang, Vida Zhang, Jessica Zhao, Yuan Jiang, Haojian Jin, Sam Kim, Evelyn Kuo, Tianshi Li, Jinping Liu, Yile Liu, Robert Zhang

Carnegie Mellon University, jasonh@cs.cmu.edu

We present the design and design rationale for the user interfaces for Privacy Enhancements for Android (PE for Android). These UIs are built around two core ideas, namely that developers should explicitly declare the purpose of why sensitive data is being used, and these permission-purpose pairs should be split by first party and third party uses. We also present a taxonomy of purposes and ways of how these ideas can be deployed in the existing Android ecosystem.

CCS CONCEPTS •Human-centered computing~Ubiquitous and mobile computing~Ubiquitous and mobile computing systems and tools •Security and privacy~Human and societal aspects of security and privacy~Privacy protections •Security and privacy~Human and societal aspects of security and privacy~Usability in security and privacy

## 1 INTRODUCTION

Privacy Enhancements for Android (PE for Android[1]) is a DARPA-funded project seeking to improve the privacy of the Android Open Source Project (AOSP). In this white paper, we focus on one specific aspect of PE for Android, namely the new privacy-focused user interfaces that our team has been designing, prototyping, and testing over the past four years of this project. Our goal is to share with other researchers, designers, and privacy advocates our user interface designs and our design rationale, which we believe can greatly improve the state of the art when it comes to improving the smartphone privacy ecosystem. Furthermore, while our focus is with Android, we believe that many of our ideas can also help with privacy for iOS and for other platforms beyond smartphones, e.g. Augmented Reality and Internet of Things.

While we have developed many features to improve privacy, there are two core ideas to our work. *The first idea is to have developers explicitly declare the purpose of why sensitive data is being used, which can then facilitate a wide range of activities across the entire smartphone ecosystem to improve privacy.* More specifically, rather than simply declaring that an app uses "location" or "contact list" (which is how Android permissions work today), developers would also be required to declare the purpose of use, selecting from a small set of pre-defined purposes. For example, developers might declare that their app uses "location for advertising" or "contact list for backup". Together, these permissions, purposes, and other related metadata, form an app policy that is embedded as a single file inside of an app, which is used to populate different privacy user interfaces as well as enforce defined behaviors.

---

[1] See more info about Privacy Enhancements for Android at https://android-privacy.org/

Presently, developers can informally convey the purpose of use through freeform text via app descriptions in the app store, in the body of privacy policies, or directly in the user interfaces of apps. However, this kind of text is unstructured and hard for automated techniques to analyze and check. Instead, we want these purposes to be machine-readable, which would make it clearer what the expected behavior of the app is and enable many new ways of checking and enforcing privacy throughout the Android ecosystem. For example, after declaring purposes and other related metadata, developers can get immediate feedback about privacy issues as they code, as well as hints to improve their code. Development teams can also more easily conduct internal audits to see what sensitive data is being collected and check that that data is being used in an appropriate manner. App stores and independent researchers can more readily apply static and dynamic analysis techniques to examine the behavior of apps, since there is more information about what the expected behavior should be. End-users can better understand why an app is requesting sensitive data, selectively allow apps access to sensitive data for specific purposes, and ultimately make better trust decisions that they are comfortable with.

*The second core idea to our work is to separate permissions and purposes into first-party use (where sensitive data is accessed by the app itself and possibly sent to the app developer's servers) and third-party (where sensitive data is accessed by the app itself, typically through third-party libraries, and potentially sent to third-party servers).* Currently, Android does not differentiate between sensitive data being accessed by an app itself versus a third-party library. However, this kind of information can offer end-users a useful perspective on the purpose of use, for example, helping them distinguish between location data being sent to the developer's own servers vs being sent to Facebook, Google, or even a third-party advertising network. This information can also give end-users hints as to whether the request for sensitive data is essential to the functionality of the app or incidental. Note that distinguishing between first-party and third-party access only describes the app executable itself, and does not say anything about onward transfer of data from first-parties to other third parties. For instance, an app might access and send data to the app developer's servers, which is then shared with other third parties.

Stepping back, we focused on these two ideas based on several constraints we imposed to guide our work. In particular, rather than proposing radical changes to the implementation of Android apps, we wanted to support some degree of backwards compatibility as well as offer a pathway for the widespread adoption of our ideas by the Android ecosystem. In addition, we wanted our approach to be easily understandable by developers and a low burden for end-users. While we also had many speculative ideas for improving Android privacy (described in more detail near the end of this document), we focused primarily on those that we felt were more reliable for designers, developers, and end-users.

These two core ideas guided a number of designs for our user interfaces for PE for Android, including:
- A taxonomy of purposes describing why sensitive data is being used by an app, which our team created over multiple iterations of card sorting exercises.
- Policy cards as a common way of presenting to end-users what permission is being requested by an app, for what purpose, and by whom.

- A "zooming interface" to cluster similar policy cards together, while also allowing end-users to see more details if desired. This kind of clustering is necessary for apps that send data to multiple third parties, so as to not overwhelm end-users with too many policy cards at once.
- Multiple user interfaces that use these policy cards, including install time (shown when a new app is installed), runtime (shown when an app requests sensitive data), configure time (a settings screen), and organizational profiles (a way for organizations to impose mandatory privacy policies for smartphones)
- An extension to Android's Quick Settings interface, allowing end-users to easily turn sensors on or off for all apps on the phone (e.g. turning microphone or location off)
- Notifications to help explain when a policy decision has been automated based on an existing setting (e.g. an app has a policy that denies access to microphone, or location is turned off for all apps on the Quick Settings interface)
- A Privacy Home Screen that is part of Android's Settings, which links to the configure time user interfaces, and also shows recommended policies, highlight highly invasive apps (in particular, spouseware), and shows visualizations of how one's sensitive data is being used and by what app

Many of these ideas are embodied in our Policy Manager, which is an extensible module in PE for Android that lets end-users configure settings and return decisions about whether an app should be allowed to access sensitive data. Note that PE for Android has a default Policy Manager, and our team developed an alternative Policy Manager with additional features to improve usability.

All of our user interface designs and code is available on our project website https://android-privacy-interfaces.github.io. Our user interfaces are also available on Figma at https://www.figma.com/file/TDJXudKC9cjWH2Ex4M2c6BAH/Brandeis-19Spring?node-id=4034%3A14601. A working draft of our app policies and our rationale is available at https://docs.google.com/document/d/1oKg3dGWzLJ7AluFGs4Z1Vr9f_Xmujn7b6F0vPd99uww/. In the remainder of this paper, we discuss purposes, these user interfaces, and the design rationale behind our user interface designs and decisions. Lastly, given that there are many new terms introduced, we also include a Glossary as an Appendix.

## 2 BACKGROUND AND RELATED WORK

### 2.1 Concerns about Smartphone Privacy

There have been growing concerns about smartphone privacy in recent years across a wide range of issues. For example, in 2011, executives from Apple and Google testified to Congress about their data collection practices with smartphones (Vega, 2011). The Federal Trade Commission has fined several companies for apps that violate consumer privacy (Federal Trade Commission 2011; 2013a; 2013b). In 2010, the Wall Street Journal published a privacy analysis of apps entitled What They Know, showing how much data apps are collecting (Thurm and Kane, 2010). More recently, the New York Times has been showcasing opinion pieces and analysis of web and smartphone privacy issues in The Privacy Project. There are also numerous news articles debating whether Facebook's app is using the microphone to spy on people (see for example (Tiffany, 2018)).

There have also been numerous research papers looking at smartphone privacy, for example people's lack of understanding of permissions (Felt et al, 2012), what leads people calling smartphone apps "creepy" (Shklovski et al, 2014), the privacy of apps with respect to children (Liu et al, 2016; Reyes et al, 2017), privacy in participatory sensing (Shilton, 2009), and many, many more. Perhaps most pragmatically, a 2015 Pew Research Center survey (Olmstead, K. and Atkinson, M.) found that 60% of people chose not to install an app when they discovered how much personal info it required, and 43% uninstalled an app after downloading it for the same reason. Thus, if we want more people to adopt smartphone apps, we need to find ways of legitimately addressing their privacy concerns.

However, privacy is challenging for a number of reasons. Hong outlines a number of these reasons (Hong, 2019), including: (a) smartphones have access to a rich and wide range of sensitive data through their sensors and data stores; (b) it is relatively easy for developers to access this sensitive data; (c) most developers have little knowledge about privacy issues, let alone what to do about them with respect to software architecture and user interface design; (d) there are few tools to help developers design, implement, and audit for privacy; (e) there are often strong incentives for companies to collect lots of data about their users; (f) it is not clear what are best practices with respect to user interface design; and (g) much of managing privacy today places a high burden on end-users, who may lack the awareness, knowledge, and motivation to protect themselves.

In short, while smartphones offer many capabilities and benefits, there are also numerous privacy problems that have been documented, and those problems are only growing.

## 2.2 Smartphone Permissions

Currently, Android requires app developers to declare up front what permissions an app will use. In older versions of Android, these permissions would be presented to end-users at install time, right before an app was installed on the smartphone. However, in more recent versions of Android, the install time user interface has been removed, with Android instead favoring runtime user interfaces that are presented at the time sensitive data is requested by an app.

A great deal of research has found several challenges with Android's permission model. For example, past studies have shown that mobile users have a poor understanding of permissions (Chin et al, 2012; Felt et al, 2012; Kelley et al, 2012). They have low attention at install time and cannot correctly understand the permissions they grant, and current permission warnings are not effective in helping users make security decisions. There is also a lack of clarity as to why sensitive data is being used by an app (Lin et al, 2012). Furthermore, once sensitive data is allowed, it can be used for multiple purposes throughout an app.

There have also been several studies examining how people make decisions about whether to install an app or not, finding that factors such as perceived utility (Egelman et al, 2013), purpose of use (Lin et al, 2012), or the price of the app (Shklovski et al, 2014) have a strong influence.

To address these and other privacy concerns, Google initiated Project Strobe as a way of managing third party access to sensitive data. In 2018, a blog post detailed some changes to permissions, limiting the granularity of access to Google Account data, as well as limiting the use cases for access to SMS, contacts, and phone permissions (Smith, 2018).

Because permissions are a fundamental part of Android (and to a lesser extent iOS), we opted to build on permissions in our work on PE for Android rather than seek a radical alternative.

## 2.3 Designing Better User Interfaces for Privacy

Currently, Android offers several user interfaces for privacy. Figure 1 shows the Android settings screens for managing permissions (left) and specifically for location (right). Starting in version 6.0, Android also requires developers to get explicit permission at runtime from users of their app to access sensitive information, presenting a user interface that is generated by the Android operating system (see Figure 2).

Note that this runtime dialog is not customizable by the programmer, and any context for the permission being requested depends on the programmer to request the permission at the right time. Apps can, if they choose, display custom interfaces to their users which provides some context for sensitive data access. These interfaces are not provided by Android, however, and so are only as useful as the app makes them. Furthermore, because the interfaces are custom-made, there is no guarantee of any consistency across similar apps. Lack of consistent explanation for data use across apps can lead to further confusion for users. Figure 3 shows two examples of custom user interfaces.

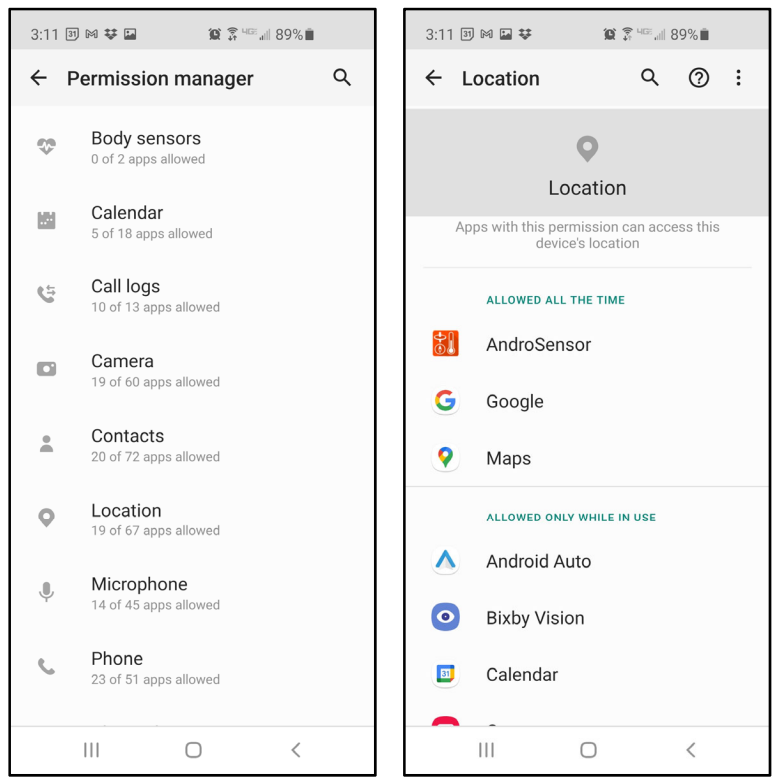

**Figure 1.** User interfaces for managing permissions in Android 10 (also known as Android Q). The left side shows the list of permissions, and the right side shows settings for location. Note that most permissions only offer "allowed" or "denied", with location offering "allowed all the time", "allowed only while in use", and "denied".

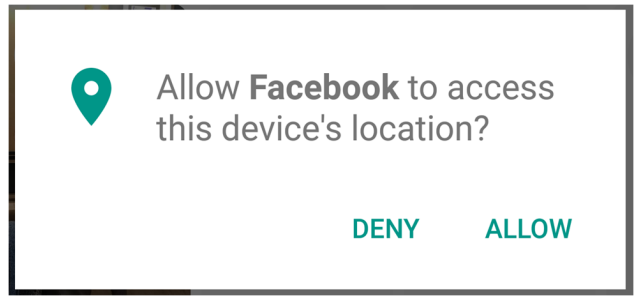

Figure 2. Android runtime user interface for denying or allowing a permission request by an app, displayed automatically by the operating system.

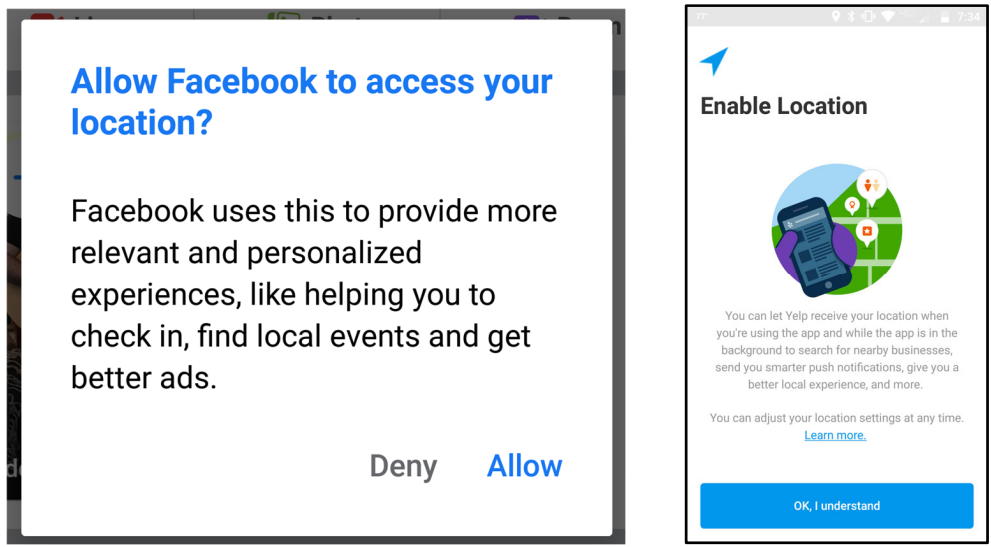

Figure 3. Two example custom user interfaces for requesting access to sensitive data.

For comparison, iOS does not have explicit permissions, but does offer several user interfaces for managing privacy and sensitive data. Figure 4 shows several iOS user interfaces, including a UI for configuring sensitive data somewhat analogous to Android's in Fig 1, and a runtime UI requesting access analogous to Android's in Fig 2 and Fig 3. iOS also allows developers to present custom purpose strings inside of runtime permission requests, though these are arbitrary text strings (Tan et al, 2014).

There has been a great deal of research looking at how to improve these privacy user interfaces on smartphones. Some examples include showing how comfortable other people were with a given permission (Lin et al, 2012), displaying just-in-time notifications in the status bar showing that sensitive data was accessed (Balebako et al, 2013; Enck et al, 2014), presenting runtime user interfaces that surface what decisions other people made (Agarwal and Hall, 2013), showing people after the fact what data was collected by an app (Balebako et al, 2013; Almuhimedi et al, 2015), allowing people to allow or block permission requests on a per-library basis rather than a per-app basis (Chitkara et al, 2017), as well as investigating how contextual factors such as location or time of day might be used to help automate decisions (Olejnik et al, 2017; Wijesekera et al, 2017).

There have also been several research systems letting people examine the privacy of apps outside of a smartphone. For example, PrivacyGrade.org analyzes apps based on purpose of use, as determined by third-party libraries used by the app. X-Ray Refine (Van Kleek et al, 2018) lets users see destination types of apps (e.g. advertising and analytics), geographic destination, and some purposes of use.

With PE for Android, our main focus was designing and implementing privacy interfaces that embodied two key ideas, namely purposes and differentiating between first- and third-party uses. As part of this work, we developed a suite of UIs for privacy, including install time, runtime, configure, notifications, quick settings, and organizational profiles.

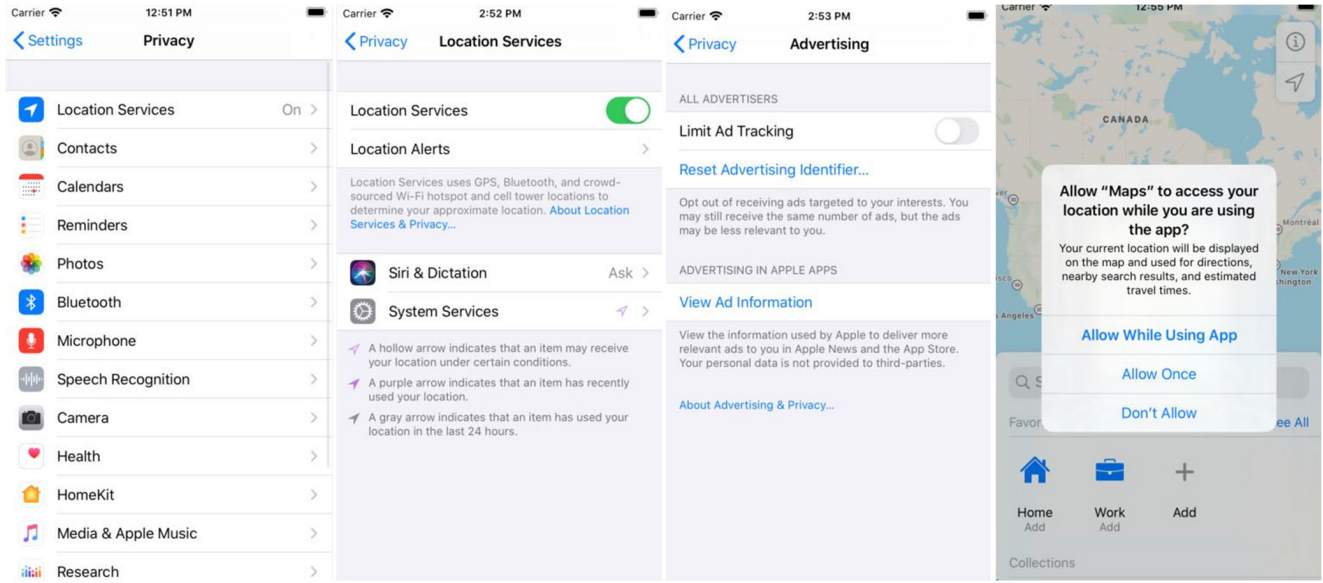

Figure 4. iOS offers several user interfaces for managing privacy, with respect to different sensors and personal data (far left), location data (middle left), advertising (middle right), and runtime user interfaces for requesting access when using an app (right).

**2.4 Purpose of Use for Sensitive Data**

A great deal of past research has found that understanding why an app uses data can have significant influence on whether people would allow that data access or not (Lin et al, 2012; Tan et al, 2014; Olejnik et al, 2017; Wijesekera et al, 2017). However, as noted earlier, this kind of information is currently only conveyed in informal ways, for example through the app description or through text strings in the app's user interfaces, if it is conveyed at all. For example, a third-party library might access sensitive data without the user or even the developers being aware at all, see for example (Agarwal and Hall, 2013; Balebako et al, 2014; Li et al, 2017).

To address this problem of understanding app behavior, one line of research has been to develop new kinds of techniques to infer these purposes. For example, PrivacyGrade (Lin et al, 2012; Lin et al, 2014) and newer versions of ProtectMyPrivacy (Agarwal and Hall, 2013) infer purposes by looking at the third-party libraries that are requesting sensitive data. WHYPER (Pandita et al, 2013), CHABADA (Gorla et al, 2014), and AutoCog (Qu et al, 2014) use the text-based app description to infer purposes of why sensitive data is used. MobiPurpose (Jin et al, 2017) examines network traffic to infer purpose of use, using features such as network payload, text from the URL path, and destination. However, these inferences are still best guesses, and as such are limited in how they can be applied. Note that these techniques can be used to help generate app privacy policies for existing Android apps so that they can be run on PE for Android. These techniques can also be used to help analyze the behavior of apps targeting the PE for Android platform to help ensure that developers are being honest about their apps' declared purposes.

In our initial work on smartphone privacy, our core idea was that we could use crowdsourcing to measure people's expectations about what data an app might use, which we found was strongly correlated with people's privacy comfort levels (Lin et al, 2012). We also found that explaining why an app was using sensitive data uniformly led to lower privacy concerns. This work was also our first connecting libraries with purposes. For example, if an app uses sensitive data because of a third-party library, that library offers a large hint as to why that data is being used (e.g. advertising or social media).

In later work (Wang et al, 2015; Wang et al, 2017), we extended this analysis to custom code (rather than just libraries), by decompiling APKs and examining the text strings that remained, which offer a large hint as to the purpose of use. For example, if we pinpoint that location data is being used by a certain part of the code, and then find strings like "photo" and "geotag" in the names of classes, methods, or variables, then a reasonable guess is that the app is using location data for geotagging purposes. To help train a machine learning classifier, we manually decompiled and inspected over a thousand apps that used location data or contact list, grouping them together based on their purpose. In practice, we found that there were only about 10 primary purposes for location data (e.g. search nearby, transportation, geotagging, reminders) and 10 purposes for contact list (e.g. backup, blacklist, email, find friends).

This current white paper presents an alternative approach to purpose of use, which is to have developers explicitly declare those purposes. However, as noted above, techniques that can help infer purposes are still useful to help in terms of generating purposes for existing apps and for checking any declared purposes to ensure that those purposes are correct.

In late 2020, Apple released their privacy nutrition labels, which describe data types an app might use and their purpose. Note that their set of purposes is much smaller than our taxonomy. For example, their purposes include Product Personalization, Third-party advertising, Developer's Marketing or Advertising, Analytics, and App Functionality. Furthermore, these privacy nutrition labels are separate from an iOS app, rather than an intrinsic part of it as is the case with PE for Android.

### 2.5 System Architectures for Smartphone Privacy

Researchers have also investigated many different system architectures to improve smartphone privacy. Note that there are a large number of research papers in this space, and our focus in this paper is more about user interfaces for privacy, and so we only list a few representative examples for each.

Some of this work has investigated modifications to the smartphone operating system. For example, TaintDroid (Enck et al, 2014) added modifications to Android to track sensitive data as it was accessed from a source and trace it to a sink. Mockdroid (Beresford et al, 2011) let users choose whether to empty data or actual data should be sent to apps when sensitive data is requested by Android apps. ProtectMyPrivacy (Agarwal and Hall, 2013) intercepted calls to sensitive data on iOS to let users choose at runtime whether to allow or not and see other people's decisions. A version of ProtectMyPrivacy for Android (Chitkara et al, 2017) let users set pre-configured policies rather than always deciding at runtime, and also allowed users to set policies on the granularity of third-party libraries rather than just apps. A field study found that this approach greatly reduced the number of decisions that users had to make.

Our work in this white paper was inspired in part by this past work. The underlying PE for Android system also introduces modifications to Android OS, in the form of a Policy Manager that lets users configure settings, record user decisions, and respond to requests for sensitive data by apps. Our work presented in this paper differs from this past work in its focus on user interfaces, especially its investigation of purposes and app policies. ProtectMyPrivacy was also developed by members of our team, and the experiences and user studies helped inform the design of our user interfaces.

## 2.6 App Analysis

There have been a large number of smartphone app analysis tools created by researchers, and we only list a few examples here. Some of these tools examine app metadata to understand what sensitive data an app might use (Pandita et al, 2013; Gorla et al, 2014; Qu et al, 2014). Others apply static analysis techniques to app binaries to understand information flow or privacy-related app behaviors (Arzt et al, 2014; Li et al, 2015; Wang et al, 2015; Pan et al, 2018). There are also tools that apply dynamic analysis techniques to understand the security of apps, information flow within an app, or what data flows out of an app and why (Burguera et al, 2011; Wei et al, 2012; Zheng et al, 2012; Rastogi et al, 2013; Avdiienko et al, 2015; Xia et al, 2015; Yang et al, 2015; Wang et al, 2017; Jin et al, 2018; Ren et al, 2018). There have also been some commercial ventures in this space. In particular AppCensus.io offers a service that analyzes apps and "detects behaviors relevant to privacy policies, laws, and regulations."

App analysis fits into our work in two different ways. The first is in generating app policies for existing apps. The main idea here is that there are a large set of existing Android apps, and for them to work with PE for Android, we need to generate app policies that can describe the sensitive data they use and the purpose of use. In our work here, we apply relatively simple static analysis techniques, looking at what third-party libraries an app uses and some of the package names to infer a likely purpose, a simpler variant of our past work in this area (Wang et al, 2015). See Section 3.3 for more details.

The second is applying these app analysis techniques to check app policies embedded in new apps that specifically target PE for Android. In this case, app stores could check these app policies for correctness and completeness. Note that we have not implemented these ideas, but rather describe how it could work in Section 7.3.

## 3 OVERVIEW OF APP POLICIES

To recap, app policies describe an app in terms of the permissions that it might use and the purpose of use, along with other related metadata.

A major question here is, how are these app policies created? There are two major cases we need to address. The first is new apps. Here, app policies are expected to be created with the help of developers and embedded inside of Android APKs in a well-known location at compile time. The second is existing apps. There are already several million Android apps in existence that do not have clear descriptions of purpose of use, and so for PE for Android and its ideas to be widely adopted, we need a way to accommodate these existing apps.

Briefly, our main idea for this case is to generate app policies for these apps by applying a suite of analysis techniques to infer the purposes of use. Given the scale of this problem, we narrowed our scope by generating app policies for just the most popular apps. We also use relatively simple methods, analyzing what libraries those apps use (an approach that we have used to good effect in past work (Lin et al, 2012; Lin et al, 2014; Chitkara et al, 2017), and as a fallback case use an app's existing permission information plus inspection of package names and class names to get an idea what data is used for. Table 1 presents a summary of these different cases. In the remainder of this section, we describe these approaches and the structure of app policies in more detail.

| Different Cases for App Policies | Description |
| --- | --- |
| New PE for Android Apps | Developers are expected to use Coconut IDE Plugin + Privacy Annotations to generate app policies. These app policies are then embedded in APK. See Section 3.1 for more details. |
| Existing Android Apps (popular apps) | App policies are pre-generated for the most popular apps. App policies are currently generated based on what third-party libraries are used as well as keywords in package names (e.g. "ads", "social", "weather"). See Section 3.2 for more details. |
| Existing Android Apps (all other apps) | Fallback case, use existing Android permission system and make clear that purpose of use is unclear. See Section 3.2 for more details. |

**Table 1.** Summary of how app policies are supported for both new apps targeting PE for Android and for existing Android apps.

### 3.1 The Structure of App Policies

The app policy itself is implemented as a JSON object. See Figure 5 for an example snippet of an app policy (see Appendix C for a fuller app policy for WhatsApp). Table 2 gives some more details about each of the fields.

App policies are used in several different parts of PE for Android. When installing an app (i.e. install time), PE for Android will read in the app policy and use that to populate data fields in our new install time user interface (see Section 5 for more details about the user interface design).

App policies are also used by our Policy Manager for PE for Android. As noted earlier, the general role of the Policy Manager is to let end-users configure what sensitive information apps can access. In our extension, we store those decisions as user policies, and return a decision to PE for Android of one of "allow", "ask", "deny" when sensitive data is requested by an app. PE for Android forwards requests for certain permissions to the Policy Manager, such as location data or contact list. PE for Android currently uses Android's set of dangerous permissions for this, see Appendix A for a full list of these permissions.

A major design decision is whether access policies should focus on access (when sensitive data is requested) or use (when that sensitive data is actually used). We chose to support access, primarily due to simplicity. That is, it is relatively easy to determine when an app is requesting sensitive data because it is a matter of checking against a known set of API calls. In contrast, it is rather hard to determine all the places that an app is using data, as it requires flow analysis. A tradeoff, however, is that it can be harder for analysis tools to check the purpose of use at the point of access.

```json
{
  "uses": "android.permission.ACCESS_FINE_LOCATION",
  "purpose": "Advertisement",
  "class": "com.mopub.mobileads.MoPubView",
  "method": "loadAd",
  "for": "Delivering relevant ads, which help to keep this app free."
}
```

**Figure 2.** Example snippet of an app policy, represented as a JSON object. App policies describe what sensitive data will be used, for what purpose (both machine-readable and human-readable), as well as where in the code the data will be accessed.

| Field Name | Description |
|---|---|
| Uses | The name of the permission, same as in the Android manifest. PE for Android focuses primarily on dangerous permissions, see Appendix A for a full list. |
| Purpose | The name of the purpose, from the taxonomy of purposes. Section 4 discusses purposes in more detail. See Appendix B for full list of purposes. This information is presented to end-users in PE for Android's user interfaces. |
| Class | The fully qualified class name in the APK that is using the sensitive data. The class and method fields are used internally by PE for Android's Policy Manager, to help disambiguate different runtime requests for sensitive data within an app. |
| Method | The name of the method in the specified class (above) that is using the sensitive data. See Class (above) for more details. Note that we currently only use the method name, for reasons of simplicity. In practice, this field should use the full method signature (including parameters), so that there is no possible ambiguity if there are multiple methods with the same name. |
| For | The purpose string, a human-readable string describing the purpose in more detail. These strings are presented to users directly in PE for Android's user interfaces. Note that we have not currently set a limit on the length of this field, but rather truncate it in the user interface if too long. |

**Table 2.** Description of the different fields in an app policy. These fields are expected to be specified by the developer, though tools can help in many cases (e.g. the Uses, Class, and Method fields).

Note that in today's version of Android, an app only needs to specify what permissions are needed. This leads to the problem of an app using the same data for multiple purposes. To address this problem, our specification of app policies require developers to also declare the class and method that is requesting the permission. This approach makes it possible for the Policy Manager to disambiguate requests for permissions, and also makes it possible for end-users to allow (for example) "location for geotagging" but deny "location for social media".

There is potential for source code to become inconsistent with app policies, for example a developer renames a class but does not update the app policy, but we expect IDE tools can automate the process of generating app policies, thus mitigating this problem. Another potential concern is that our approach assumes that there will be at most one purpose associated with each class and method pair, for example that an app does not use the same class plus method pair to request location for geotagging and for social media purposes. A related issue is that our approach means that apps should not request sensitive data in one method for a given purpose, and then use it for other purposes in other methods. For example, in some of our analyses of apps, we have seen a few apps that pre-fetch location data in one method, and then make that data available to the rest of the app for multiple purposes (e.g. advertising, social media, maps). While we do not currently address these related problems, we believe IDE tools can push developers towards software engineering practices that are easier to analyze and handle without requiring major changes to source code. Furthermore, we believe that app stores can include static and dynamic analysis techniques that can detect these cases and require developers to conform to specific coding practices.

Another potential issue is, what happens if there are errors in the app policy? In our current implementation, we assume that app policies are correct, in that app developers have honestly and correctly specified their app's behaviors. We note that there are two possible failure modes here, namely that (a) app developers may incorrectly specify behaviors, or that (b) app developers are malicious and deliberately trying to mislead app stores and end-users. Our working assumption is that tools and internal audits can help prevent the former case, and that app analysis tools can detect the latter case when apps are uploaded to app stores (see Related Work and Discussion for examples). It is also possible for PE for Android smartphones to randomly check behaviors of apps to ensure proper behavior, to balance additional runtime overhead with safety. For example, when an app requests sensitive data, a future version of PE for Android might apply some heuristics or machine learning techniques to the stack trace to see if the inferred purpose makes sense. See (Wang et al, 2015; Chitkara et al, 2017; Wang et al, 2017) for examples of modifications to Android that do this kind of runtime checking.

This current implementation of app policies is a subset of our broader ideas for what app policies could be used for. For a more detailed discussion, please see our working draft of app policies and our rationale at https://docs.google.com/document/d/1oKg3dGWzLJ7AluFGs4Z1Vr9f_Xmujn7b6F0vPd99uww/.

### 3.2 Supporting App Policies in New Apps

To support new apps in creating app policies, we have created a custom Android Studio IDE plug-in called Coconut (Li et al, 2017). In a nutshell, Coconut looks for API calls requesting sensitive data (ingress) and when sensitive data is stored or is sent over the network (egress), and then requires developers to add privacy annotations (using Java's annotations feature) to their source code describing the purpose of use and other metadata about how the requested data will be used (see Figure 6). When compiling an app, these annotations can be directly translated into an app policy and added to an APK. More details about Coconut, as well as the source code, are available at https://coconut-ide.github.io. Note that there are two versions of Coconut, one for Android (which does not generate app policies) and one for PE for Android (which does).

Presently, for new PE for Android apps, we store app policies in an APK under `/assets/odp/policy.json`. Here, odp stands for Off-Device Policy, which is an alternative name our team sometimes uses for "app policy" (since the app policy helps govern what data might leave the device).

A question here is, why would developers adopt Coconut and privacy annotations? The main reason is that compliance imposes new kinds of pain points. Here, there are two major kinds of compliance. The first is compliance with new laws and regulations, such as GDPR and CCPA. The second is compliance with app store policies, such as Google Play. That is, app stores might require new apps to have app policies, or incentivize developers that use app policies. Some examples include improving the search rankings of apps with app policies, making the privacy info more prominent on the download pages for individual apps, and making it easier to compare apps based on privacy. Combined, these forces could strongly motivate developers to adopt these new kinds of app policies.

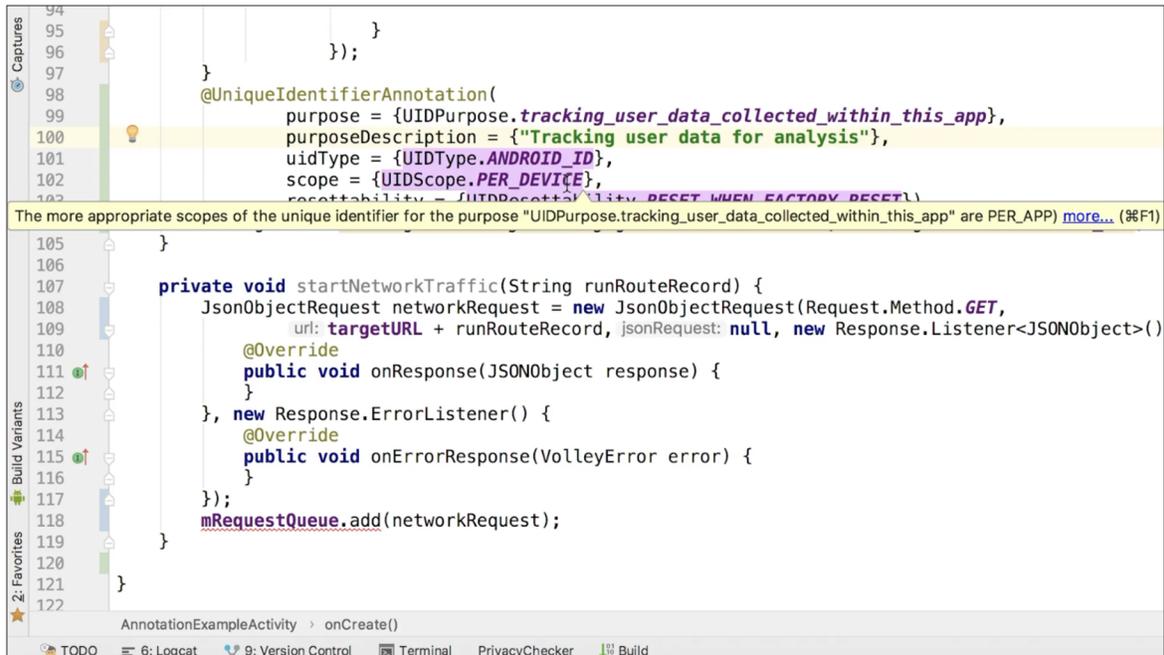

**Figure 3.** A screenshot of the Coconut IDE plugin. Coconut requires developers to add privacy annotations describing how sensitive data will be used (see @UniqueIdentifierAnnotation in the example above). Coconut also gives developers feedback about potential privacy problems as they code. One such example can be seen in the tooltip above.

### 3.3 Supporting App Policies in Existing Apps

While we have a clear path for how to generate app policies for new apps, the large base of existing Android apps lack explicit information about purposes. We believe that app stores or other trusted parties can generate reasonably good app policies for these existing apps using a wide variety of heuristics, thus giving users some level of insight about a given app's behaviors and offering a pathway for widespread adoption of our ideas with PE for Android.

Our initial approach uses two relatively simple approaches. The first is to just look at what third-party libraries a given app uses, as these libraries offer a significant hint as to the purpose of use for sensitive data. As a concrete example, if our analysis indicates that an app uses a known advertising library, and we know from prior analysis that this advertising library uses location data, we make the simple inference that this app uses location for advertising. More specifically, given an app, we inspect what third-party libraries the app uses, and then for each third-party library, we use a lookup table as to what sensitive data that library uses and the purpose of use. This lookup table is populated based on our manual inspection of those libraries and associated web sites.

This approach works well for several reasons. First, it is very simple, making it easy to implement and scale up to all existing apps. Second, there is a long tail of third-party libraries, meaning that we only need to cover the most popular libraries to be able to start generating useful app policies (Lin et al, 2012; Lin et al, 2014; Chitkara et al, 2017). Third, in our team's past work in analyzing a large number of stack traces used by apps (Chitkara et al, 2017), we found that about 40% of apps only use sensitive data because of third-party libraries. This means that for many apps, this approach is also complete. Again, we note that this is simply our initial approach, and it can be extended with more sophisticated approaches later on (see our discussion in Section 7.3).

Our second approach is to look at Android API calls requesting sensitive information inside of an app. However, note that in this case, we do not have a clear purpose. As a heuristic, we examine the package name of the class making the API call and look for words like "ads", "social", and "weather" to infer a likely purpose. If we still cannot infer a purpose, we use a generic fallback purpose called "For Other Features".

As a proof of concept, we have generated app policies for some of the most popular apps, which are included in our user interface release of PE for Android. Note that this is meant to be a temporary solution just for initial deployment. We chose this approach rather than having a server for retrieving pre-generated app policies for four reasons: (a) to avoid possible network and timing response restrictions in Android with respect to installing apps, (b) to avoid an additional network delays to retrieve an app policy if an app has already been downloaded and installation has started, (c) to minimize dependencies on our remote servers that might not be fully supported in the long term, and (d) to avoid privacy concerns, specifically that our servers would record what apps people are installing.

As such, for our current implementation of PE for Android, we have two cases to handle for existing apps (see Table 1). There are apps that we have analyzed and have app policies already generated, and those we have not analyzed. The former case is easier, in that we need to simply convey to end-users that the information they are seeing are best guesses based on automated analysis. The latter case is harder, in that we have to fall back on permissions and cannot make the purpose of those permissions clear.

For an actual wide-scale deployment, our approach of including pre-generated app policies would likely be infeasible, due to apps changing what libraries are included over time as well as the sheer number of apps that exist. A back of the envelope estimate is 3 million Android apps multiplied by about 1k per app policy leads to 3GB of app policies (uncompressed). Instead, it makes more sense to serve these app policies via app stores themselves and modify the app store app (e.g. the Google Play app) so that when people download an app the corresponding app policy is simultaneously downloaded behind the scenes.

## 4 TAXONOMY OF PURPOSES

Our core idea of having developers explicitly declare purposes means that there needs to be a canonical taxonomy of purposes that developers can choose from. Appendix B lists our current taxonomy of purposes. Here, we briefly describe our process for creating this taxonomy. As we noted in Section 2.4, Apple's privacy nutrition labels require developers to declare purposes, though they offer a smaller and coarser grained set of purposes and these purposes are separate from the actual app.

Our idea of purposes came about from the observation that Android apps (and iOS) offered user interfaces that conveyed what data was being used, but not why. In multiple informal conversations with others, this lack of a clear purpose seemed to be a major source of discomfort with respect to privacy. In some of our early work, we found that offering even simple explanations reduced people's concerns about apps using sensitive data (Lin et al, 2012). However, past work has only focused on inferring purposes, e.g. by analyzing what third-party libraries an app used (Lin et al, 2012; Lin et al, 2014), decompiling apps and analyzing the remaining text strings (Wang et al, 2015; Wang et al, 2017), or applying NLP algorithms to analyze the text description of an app (Pandita et al, 2013; Gorla et al, 2014; Qu et al, 2014).

We briefly considered having developer-defined purposes, but this would lead to a lack of consistency, would not facilitate app analysis since there would not be clear semantics, and make things harder to explain to end-users. Instead, we opted for defining a set of purposes. The challenge then is to develop a taxonomy of purposes that is mostly complete (in terms of covering the vast majority of uses of sensitive data), appropriately scoped in granularity (a given purpose would not be too broad, or too narrow and applicable for just a few apps), and easily understandable, both for developers and end-users.

Towards this end, we started with an initial set of purposes taken from our past work on text mining of Android apps (Wang et al, 2015). The core idea is that, unless obfuscated, compiled Android apps retain the class names and method names that developers give them, which can give a big hint as to the purpose of use of sensitive data. For example, if analysis indicates that an app uses location data, and the names of the classes and methods that use that location data contain words like "photo" and "tag", one can guess that the app is likely using location data for "geotagging". After manually inspecting several hundred apps, we created a set of 10 purposes for each of location data and for contact list data (for a total of 20 purposes), and created a machine learning classifier that could take an app and classify the purpose, with an accuracy of 85% for location and 94% for contact list.

Given this starting point, we expanded the taxonomy to cover a wider range of permissions (see Appendix A). We drew on other lists of purposes that researchers have proposed in their work (Han et al, 2012; Lin et al, 2012; Enck et al, 2014; Wang et al, 2015; Martin et al, 2016; van Kleek et al, 2017), though much of this past work has tended to be narrow or too coarse-grained. For example, van Kleek et al (van Kleek et al, 2017) has as purposes core/non-core functionality as well as marketing. As such, we believe that our proposed set of purposes is the most comprehensive set that has been proposed to date.

The full details of how we developed this taxonomy is described in our paper on the MobiPurpose system (Jin et al, 2018), which takes network traffic from an app and infers a purpose based on the key-value pairs, the URL path, and external information about the destination domain. At a high level, the taxonomy was created with the help of 12 different computer science undergraduate, master's, and PhD students, with backgrounds in mobile app development and UX design. We iterated on the taxonomy using card sorting over four independent sessions, in total manually labeling 1059 network traffic requests that we gathered from smartphone apps using a custom proxy and a custom monkey that we developed. In each iteration, participants first independently created their own taxonomy and then discussed with other participants to reach an agreement. The converged taxonomy would act as the starting point of the next iteration.

Again, Appendix B lists all of the purposes our team has enumerated. We believe that this set of purposes can cover the vast majority of use cases for today's smartphone apps. However, what happens if a purpose is not listed, for example if there is a brand new category of apps? One approach is to periodically update the taxonomy with new versions, once there are enough apps to justify a new purpose. While this approach might be slow to change, it does mean that user interfaces will be more consistent and make it easier to do data mining and other kinds of app analysis. Another approach is to allow developer-defined purposes in a limited fashion. For instance, RFC822 proposed using X- as a convention to denote experimental headers or commands. Some examples for email include X-SpamCatcher-Score or X-Originating-IP. Developers might use this approach to declare some new purposes, with the caveat that their app would be subjected to more scrutiny when uploading to an app store and having more warnings in PE for Android user interfaces, thus disincentivizing developers from declaring new purposes wholesale.

## 5 OUR USER INTERFACES FOR END-USER PRIVACY IN PE FOR ANDROID

In this section, we provide an overview of the privacy-oriented user interfaces our team has developed. As noted in the Introduction, we aimed for a degree of backwards compatibility in the design of our user interfaces, in that we strived to conform with Material Design where possible, and did not aim for radically new interactions or interface designs. We also wanted these user interfaces to be easily understandable by people already familiar with today's smartphones. Lastly, we wanted to minimize burden on end-users, in terms of reducing the number of notifications and other interruptions they might get, and reducing the number of decisions they have to make at configure time (via the Policy Manager, see Section 5.6), install time (see Section 5.7), and run time (see Section 5.8).

Appendix D shows specific interaction flows for important sequences, and Appendix E offers a site map. Our user interface designs are available on Figma at:

https://www.figma.com/file/TDJXudKC9cjWH2Ex4M2c6BAH/Brandeis-19Spring?node-id=4034%3A14601

### 5.1 Policy Cards and Zooming UIs

In PE for Android's smartphone user interfaces, purposes are presented to end-users in the form of *policy cards* which convey what data is being used (essentially the permission as it exists today), why (the purpose), and by whom (whether it is a third party or the app itself that is requesting the data, to help convey to end-users where the data might go). See Figure 7 for example policy cards, showing an App-Specific Setting (left) and Global Settings (middle and right). Policy cards also allow end-users to see and configure settings for a (permission, purpose, destination) tuple, in the form of *allow*, *deny*, or *ask*. Here, "allow" means always allow the data request to happen, "deny" means always disallow, and "ask" means that the decision to allow or deny will be deferred to the user via the runtime UI (presented when an app actually makes the request to access sensitive data, see Section 5.8).

By default, all settings are "ask", and end-users can choose to change a policy to "allow" or "deny". Our rationale for this default is that *on* would be overly permissive and lead to a lot of sensitive data being exposed, and that *off* would make many apps not work and leave users wondering why. As such, *ask* seemed to be the only default that made sense.

Note that one thing missing in this user interface is the purpose string from the For clause from the app policy (see Figure 5 and Table 2). We did not include it primarily due to reasons of time and prioritizing other parts of the design and prototype. Ideally, the purpose string would also be shown on each policy card, to give developers a short string explaining their data use.

There are two kinds of user policies and corresponding user interfaces, *app-specific* and *global*. App-specific Settings apply to just a single app, whereas Global Settings are ones that apply to all apps installed on the smartphone.

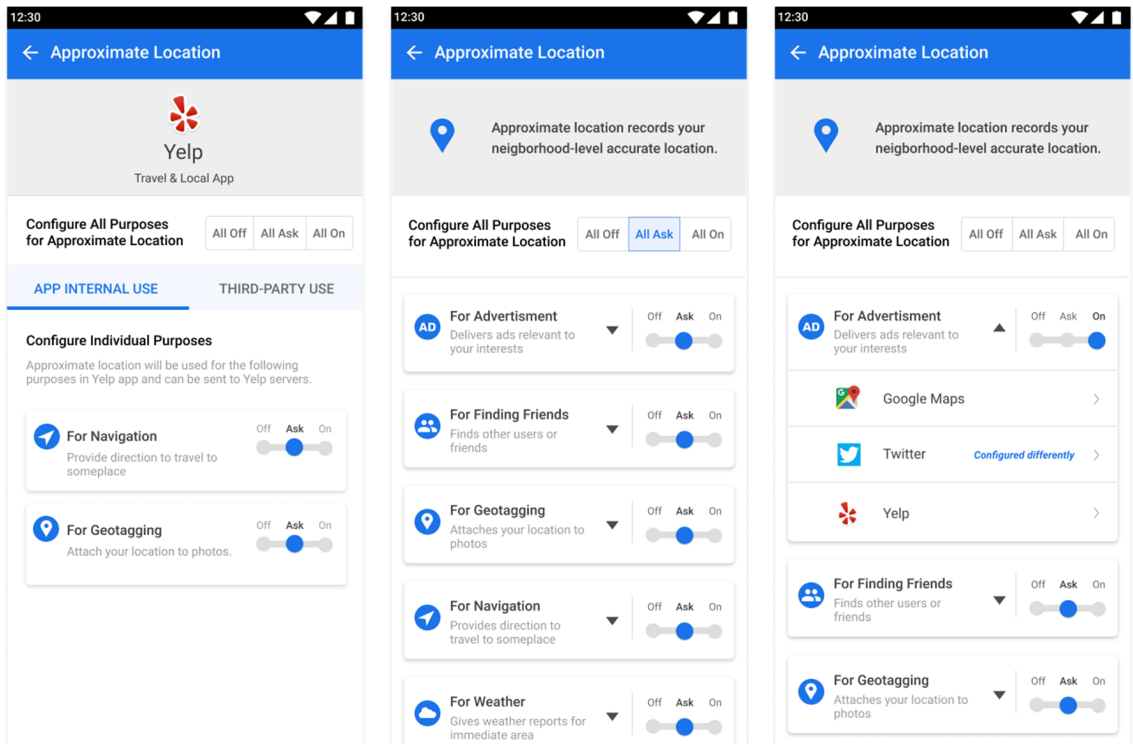

**Figure 4.** Examples of policy cards. The left screenshot shows an App-specific Setting screen containing two policy cards for the Yelp app. The middle screenshot shows five policy cards for the Global Setting for approximate location. The right screenshot shows a zoomed in view of Approximate Location for advertisements, showing policies for specific third-party destinations. These policy cards are generated directly from information in the app policies.

For App-specific Settings (see Figure 7 left), we separated the interface into settings for App Internal Use and for Third-Party Use. We originally had a single screen showing all policies, with some text to separate between these two, but informal user studies found that people were confused as to why some settings were showing up more than once (for example, an app might use Location for App Internal Use as well as for Third-Party Use). We opted to have separate tabs to differentiate these, which we felt would help clarify settings, though it also has the downside of taking an extra step to see the other settings and possibly inadvertently hiding those settings. See Figures 18 and 19 in Appendix D for details on the full interaction flow.

For Global Settings, users can select a single permission and then see and configure all of the possible associated purposes (see Figure 7 middle). A user can also drill down and then see and configure details about individual apps using that permission-purpose pair. See Figures 20 and 21 in Appendix D for details on the full interaction flow for Global Settings.

For both App-specific Settings and Global Settings, we considered allowing people to configure policies for individual third-party libraries as well, as our team did in past work as described in Section 2.5. However, we felt that this would be too complex for end-users to understand. Instead, we designed the user interface so that they could choose to ask, allow, or deny third-party destinations, e.g. Google Maps, Twitter, or Yelp (see Figure 7 right). We chose not to make this feature highly prominent, though, given that informal user testing found that this was not a highly desired feature. Another design issue here is, should the third-party name be shown, a category describing it (e.g. "Advertising Network" or "Social Network"), or both? For simplicity, we went with the third-party name itself, though there are many third-party names that people will not be familiar with, e.g. "AdMob" and "Millennial Media" (both advertising networks).

In earlier iterations, we tried grouping policy cards into "common" and "uncommon", based on how frequent these permissions and purposes were for a given app category. For example, Location for Navigation is highly common for "Maps & Navigation", but Contact List for any purpose is not. However, this design did not work very well in informal user tests, as users either missed this or were confused as to how this was determined. We instead opted to organize based on App Internal Use and Third-Party Use, which worked better in surveys we ran on MTurk as well as in-lab studies evaluating understandability of permissions and purposes.

In earlier iterations, we also just displayed the names of purposes. To clarify things, we added "For" in front of each purpose. For example, instead of "Advertisement" or "Finding Friends", we labeled these as "For Advertisement" or "For Finding Friends".

We also group similar policy cards together in what we call a *zooming user interface*. Figure 7 (middle and right) shows an example of this grouping, in this case based on destination. Zooming allows end-users to see an overview of what an app will do, and drill down to see more information if desired. This approach was inspired by a design guideline in information visualization (Shneiderman, 1996): "overview first, zoom and filter, and details on demand." That is, rather than overwhelming an end-user with every possible detail, let them see a short overview of what sensitive data the app will use. Currently, we only group based on destination, though it is likely for complex apps that further kinds of grouping may be needed.

Policy cards are shown in PE for Android at *install time* (when an app is being installed) and at *configure time* (when an end-user goes to the settings screen to manage apps. These interfaces allow end-users to manage their *user policies*, and represent discretionary control over how one's data is used. To clarify, app policies are created by developers and describe the expected behavior of an app with respect to sensitive data, while user policies are decided on by end-users and specify what behaviors are or are not allowed.

App Settings and Global Settings are user settings and thus are enforced with the lowest priority in our policy enforcement algorithm. This means they are overridden by any active policy profiles, and quick settings. See Section 5.5 for more details.

However, there may still be conflicts between an App Setting and a Global Setting, as they are at the same priority level. For example, a person might have a Global Setting that says "microphone for all purposes" is denied, but an App-specific Setting for an emergency response app that says "microphone for emergencies" is allowed.

One possible approach would be to avoid conflicts, for example having a default of deny and then letting users allow things (or conversely, having a default of allow and then letting users only deny). However, this would not work for our case, as changing Android's default to allow may cause a lot of privacy problems, and a default of deny could break a lot of existing functionality. Another approach used by many access control interfaces is to have the most fine-grained setting win. However, past work in the context of file access control (Reeder et al, 2008) has shown the difficulties involved with having multiple layers of settings, essentially forcing end-users to have to debug policies.

Instead, we were inspired by past work (Reeder et al, 2001) that used *most-recent-setting wins*. That is, settings would be time ordered, with settings checked from newest to oldest. We felt that this approach would be easier to understand, and would also make it easier for people to quickly allow or deny things (rather than having to hunt for the right global or app-specific policy). A more detailed discussion of the priority for how policies are resolved is presented in Section 5.5.

### 5.2 Organizational Profiles

*Organizational profiles* offer support for mandatory policies by organizations, such as one's employer. For example, the US Navy might want personnel to not disclose location data while on deployment, so they might have a policy that "location for social media" will always be denied. Once an organizational profile is enabled, its settings are enforced with the highest priority. It overrides any existing policy a user has configured, including quick settings.

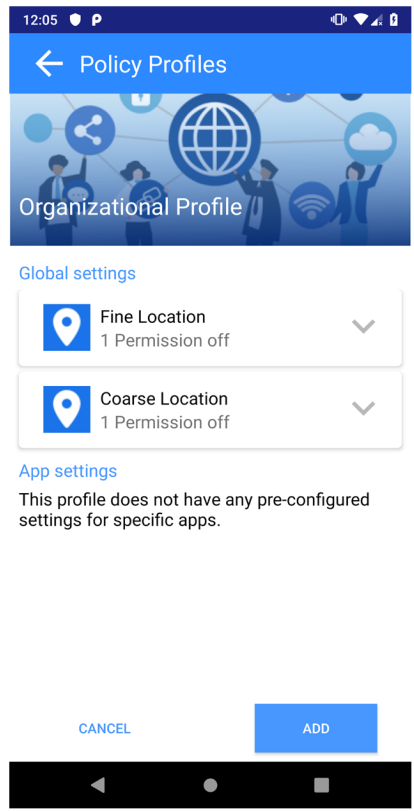

**Figure 5.** An example of an Organizational Profile for allowing organizations to set privacy policies for smartphones. Organizational Profiles have the highest priority for enforcement.

Note that these organizational policies are intended to be read only, and not able to be configured by end-users, since being able to do so would defeat the purpose. That is, organizational policies are meant to be relatively easy to add but hard to remove, to prevent circumvention by end-users.

Figure 8 shows an example user interface for an organizational profile. It displays policy cards in a manner similar to Figure 7. Organizational profiles are currently only a prototype in the policy manager. We imagine that organizational profiles could be added by scanning a QR code or directing the Policy Manager to a URL.

### 5.3 Quick Settings

Android already has a Quick Settings user interface that can be accessed by swiping down from the top of the screen twice (the first time to open the notification drawer, the second time to expand the quick settings). Quick Settings lets end-users quickly toggle common settings, such as Wi-Fi, Airplane mode, and brightness, as well as location data. We extended Android's existing Quick Settings to offer end-users a fast and easy way to turn off other sensor data for the entire smartphone, for example microphone or camera (see Figure 9). As noted by Lederer et al (Lederer et al, 2014), it can be beneficial to also have simple coarse-grained controls for privacy.

We designed Quick Settings to be enforced with the second highest priority, behind organizational profiles. In our current prototype, we chose to only support camera, location, and microphone for simplicity. We considered other sensors, e.g. humidity and temperature, but these seemed less sensitive and also less likely for people to turn off. We also considered other data types, e.g. contact list or call log, though these also seemed less likely to be used.

Note that a smartphone might have an organizational profile setting that dictates that a certain sensor must always be turned on (or off), in which case the end-user would not be able to toggle the quick setting for that sensor.

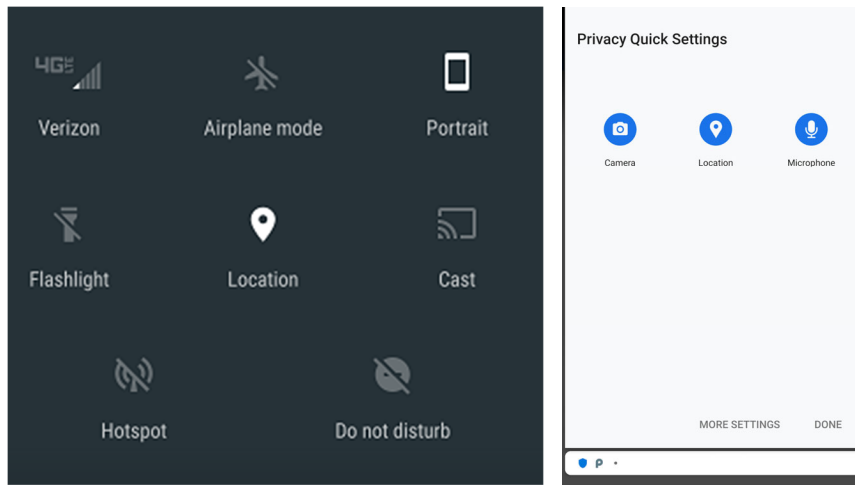

**Figure 6.** An example of Quick Settings in a previous version of Android (left), and an example of our extensions for privacy (right). In our extension, a user could swipe from the right to see more settings, one of which would be for privacy.

### 5.4 Notifications of Automated Decisions Based on Policies and for Recommendations

While the combination of organizational policies, quick settings, and user policies (both Global Settings and App-specific Settings) can help automate policy decisions, we still need a way of informing end-users of these decisions when actually running an app. That is, end-users need a way to understand what just happened and why, e.g. that camera functionality was requested and denied for Instagram because camera is turned off in Quick Settings, or location functionality was requested and denied for Blackjack game because user policy is set to deny. Given that end-users may also want to change settings, it makes sense to offer end-users an easy way of quickly getting to the right user interface.

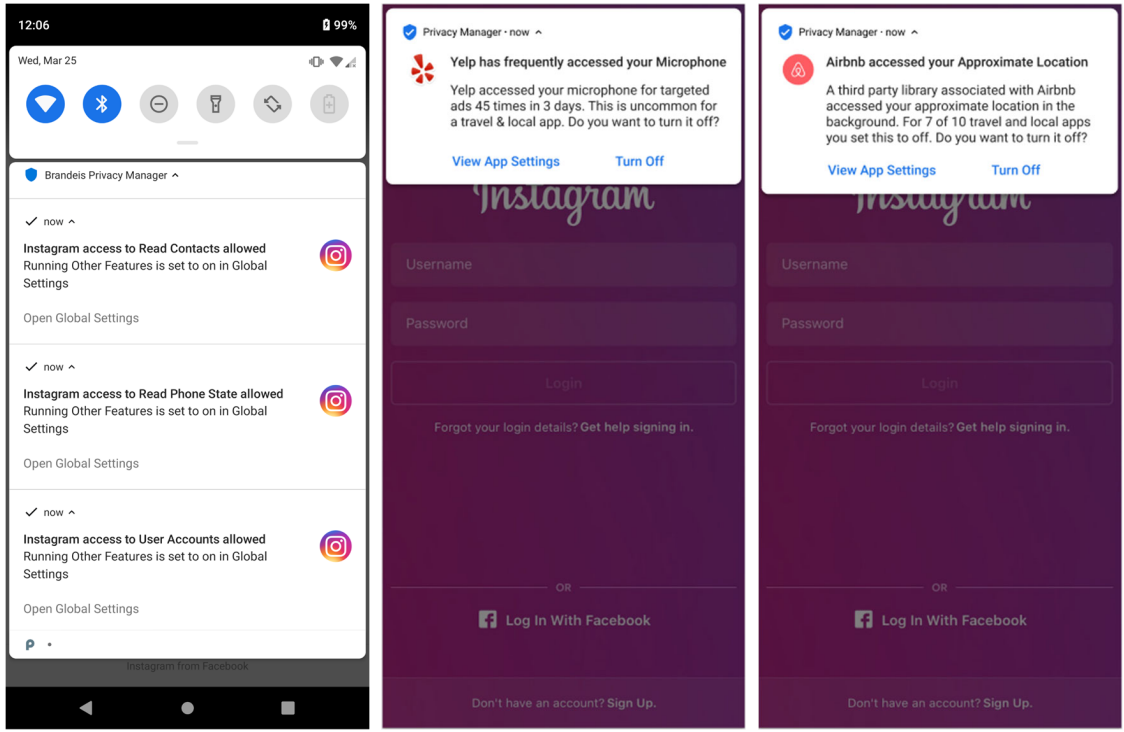

**Figure 7.** Examples of privacy notifications. Left shows a series of silent notifications informing the end-user about recent activity from a given app, in this case Instagram. Middle and right show mockups of notifications for privacy recommendations.

We offer users a log of these accesses in the Policy Manager. However, we do not expect end-users to go to the Policy Manager very often (similar to how people do not often go to the Settings screen of their smartphone). Instead, our design includes notifications that inform end-users of "allow" and "deny" automated policy decisions, including which setting led to the automated decision (see Figure 10 left). Furthermore, we designed all notifications to be silent and not alert users in any special way, since we felt that this would be a common case and numerous interruptions would quickly get annoying. Ideally, these notifications would also phase themselves out over time if no action is taken, so as not to flood the user with too much information. An alternative design might actively inform end-users of "deny" decisions periodically or if the app has not been used for a while, since people may be confused as to why the app is not working as expected.

Note that even silent alerts need to be limited in some fashion, since continuous alerts would quickly get annoying. For example, suppose for some reason a person has turned off Location data for a map app. It does not make sense to generate a new notification every time the user's location is updated. As of this writing we have not fully fleshed out such a design, but it may be the case that these notifications can be limited to once every time the app is run or once every few minutes.

We also designed these notifications to offer a quick entry point into the relevant user interface, so that they could quickly adjust their settings if desired, rather than having to hunt for the relevant screen. For example, in Figure 10 left, the notifications show that Instagram was recently allowed access to Read Contacts, and clicking on this notification would take the user to the relevant Global Settings that allowed this. An alternative design would be to take users to the App Specific settings for this app, since with our policy of most-recent-setting wins (see Section 5.1) an end-user could also set a relevant policy there.

We also designed these notifications as an entry point for *privacy recommendations*. Clearer options as to what they could do about privacy was something participants have requested since our earliest prototypes at the beginning of this project. One design option is to put them in the home screen of the Policy Manager. We did do this, but as stated above, we felt that people would rarely directly go to the Policy Manager. We also felt that occasional notifications would be a good way to surface relevant recommendations. Note that this kind of notification is a mechanism, and can be separated from how privacy recommendations are generated.

Our current design for recommendations are based on past work by our team looking at nudges based on frequency of use (Almuhimedi et al, 2015). For example, Figure 10 middle shows a notification that Yelp has frequently accessed the microphone and that this is unusual for an app in its app category (this is a mocked up example). This kind of notification gives people a better sense as to whether an access is normal or not, and also makes it easy for people to take action, in this case View App Settings or Turn Off. Figure 10 right shows a similar kind of notification for third-party libraries.

We also intend for these notifications to be used to detect spouseware running on the smartphone (spyware apps explicitly designed to spy on intimate partners). We did not implement the design or any detection software in our prototype, we expected that a simple way to start is to use a blacklist of the fully qualified names of well-known spouseware apps.

### 5.5 Flowchart for Resolving Policies

The combination of organizational policies, quick settings, and user policies also leads to another important design issue here, namely the priority in which these different settings are handled. Figure 11 presents a flowchart showing how these different settings are resolved. Roughly, the priority is as follows:
- Check Organizational Settings first
- Check Quick Settings next
- Check App-Specific Settings and Global Settings
- Ask the user

More specifically, when a policy decision needs to be made, organizational policies are checked first, since those are intended to be mandatory. If an "allow" or "deny" decision is made at this tier, then an appropriate notification is sent out letting end-users see that a policy decision was made due to an organizational policy. Otherwise, Quick Settings are checked next, which let end-users turn sensors on or off for the entire smartphone. If a "deny" decision is made at this tier, then a notification is sent out informing users that a decision was made due to a setting in Quick Settings. Otherwise, the user policies are checked, with Global Settings and App-specific Settings handled using most-recent-setting wins, along with an appropriate notification if a decision is made at this tier. Otherwise, the default decision is "ask" and the user will see the runtime user interface and will have to make a choice there.

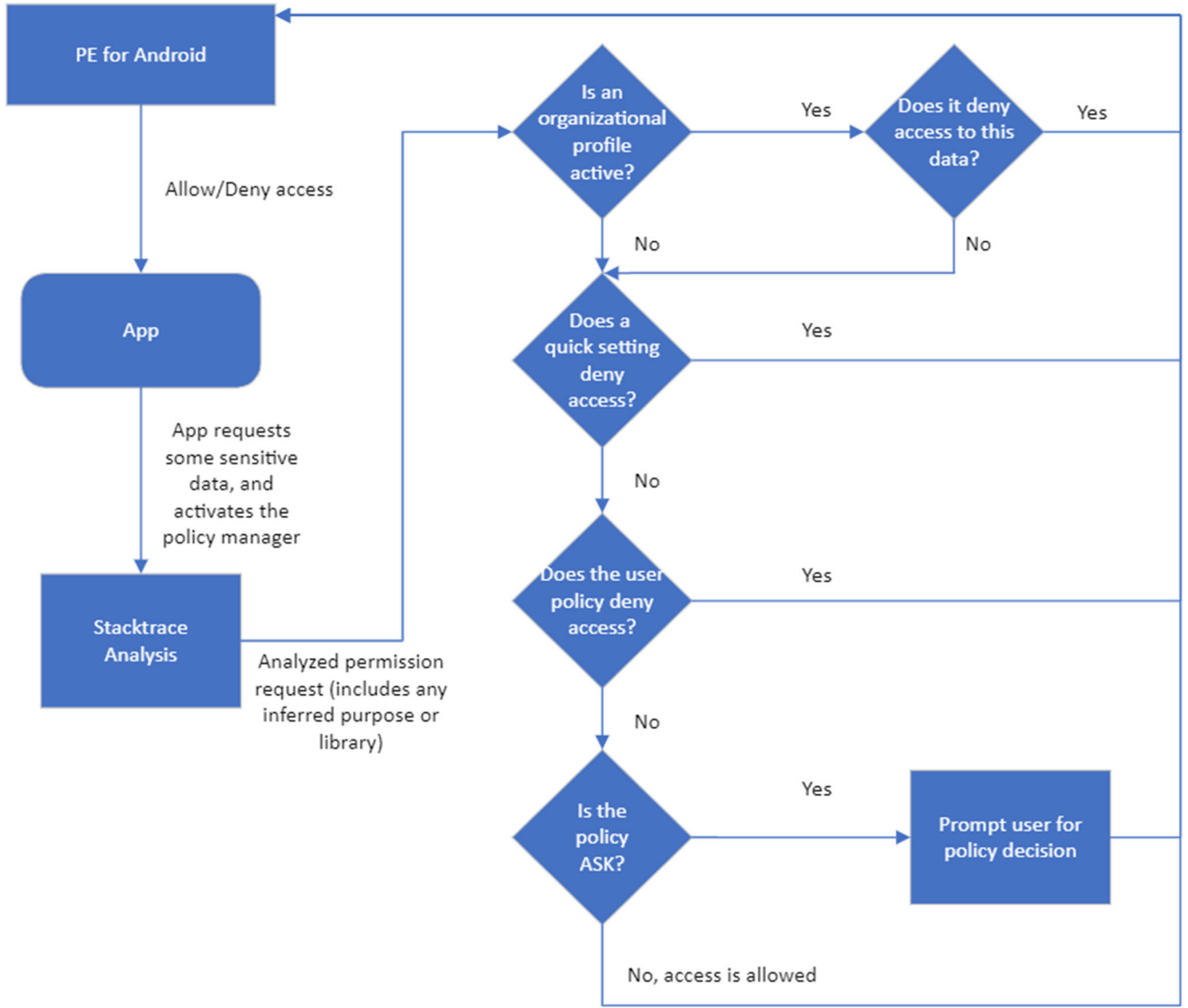

**Figure 8.** Flowchart showing the priority for how policies are resolved.

## 5.6 Policy Manager Home Screen

The Policy Manager Home Screen is the first screen a person sees when starting the Policy Manager (see Figure 12). The home screen makes it easy to see summaries of how apps have used sensitive data, offers easy access to app settings and global settings, and see recommendations.

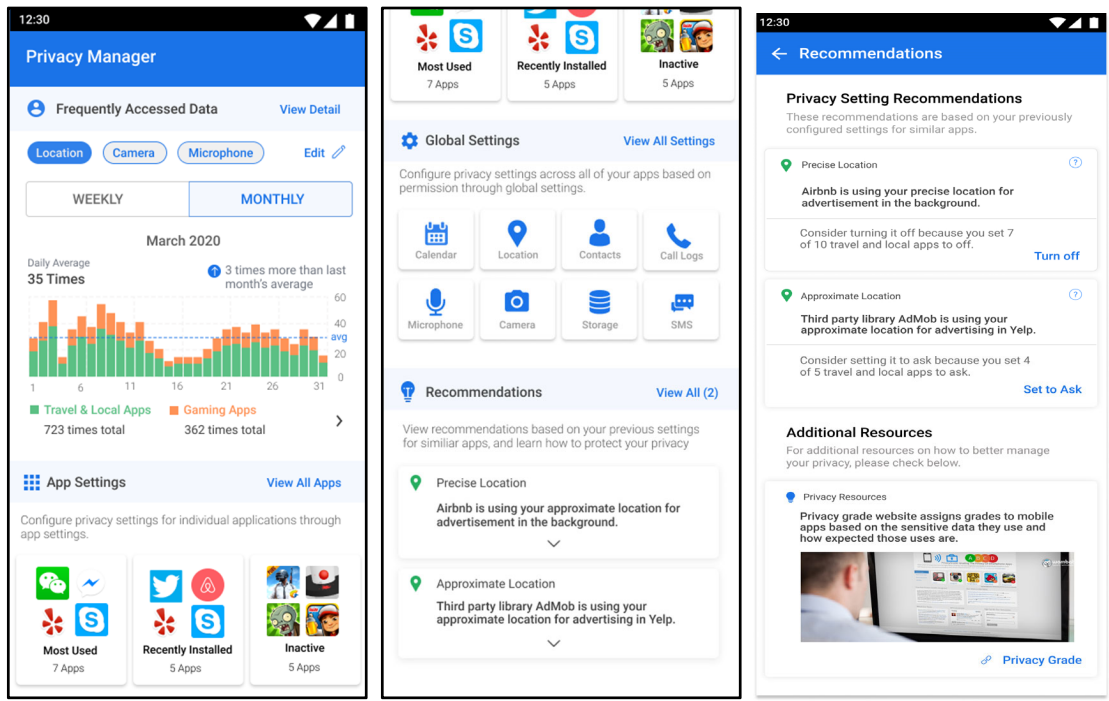

**Figure 9.** The middle and left show the Policy Manager Home Screen, split into two parts due to its length. The home screen shows a summary of how data has been used by apps, offers fast access to app settings and global settings, and presents some (mockup) privacy recommendations. On the right is the screen for Recommendations (mockup).

At the top is Frequently Accessed Data, which is intended as a space for different kinds of visualizations that convey how often different kinds of sensitive data types are being used. Users can customize what is shown from a set of data types and across different time scales.

The user can also configure settings for individual apps via App Settings. App Settings groups apps together in several different ways that we felt would be useful, e.g. Most Used apps, Recently Installed apps, and Inactive apps.

Global Settings lets users configure settings for a specific permission across all apps. We included what we felt were the most common and most sensitive data types, including Calendar, Location, Contacts, Call Logs, and more.

Lastly, the Home Screen shows recommendations for privacy settings. Here, we did not specify the exact mechanism for how recommendations are generated, instead focusing on offering a space for these. These privacy recommendations are the same as described earlier for notifications in Section 5.5. Some possibilities for privacy recommendations include simple heuristics, crowdsourcing, recommender systems based on what others specify, and using a blacklist of known spouseware apps. Note that this feature was designed but is not currently implemented in the Policy Manager.

## 5.7 Install Time User Interface

The install time user interface is shown to users right before installing an app. Previous versions of Android included an install time user interface, though later versions have dropped it in favor of runtime user interfaces. iOS also does not include an install time user interface. Apple now provides privacy nutrition labels that summarize some of the privacy-related behaviors of an app, though users have to take specific action to view those in the app store.

We chose to retain an install time user interface because we felt that our new purpose information would help users better understand how an app might use sensitive data. Furthermore, we also gave users the option to configure an app's privacy settings before installing. Lastly, some of the settings might be impacted by an organizational profile, so we felt that it was important to let users know that some functionality might not work and why.

Figure 13 presents an older design which organized policy cards by uncommon / common. The install time interface first reads in the app policy, and then sees if there are any global settings that might lock any settings as well. Figure 13 shows an example in the first policy card, where Coarse Location for Geotagging is set to Off and greyed out by DARPA.

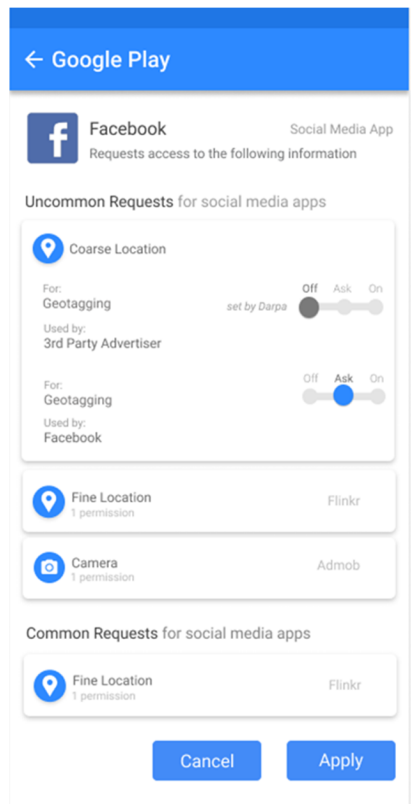

**Figure 10.** An older install time user interface for PE for Android, also showing the effects of an organizational profile.

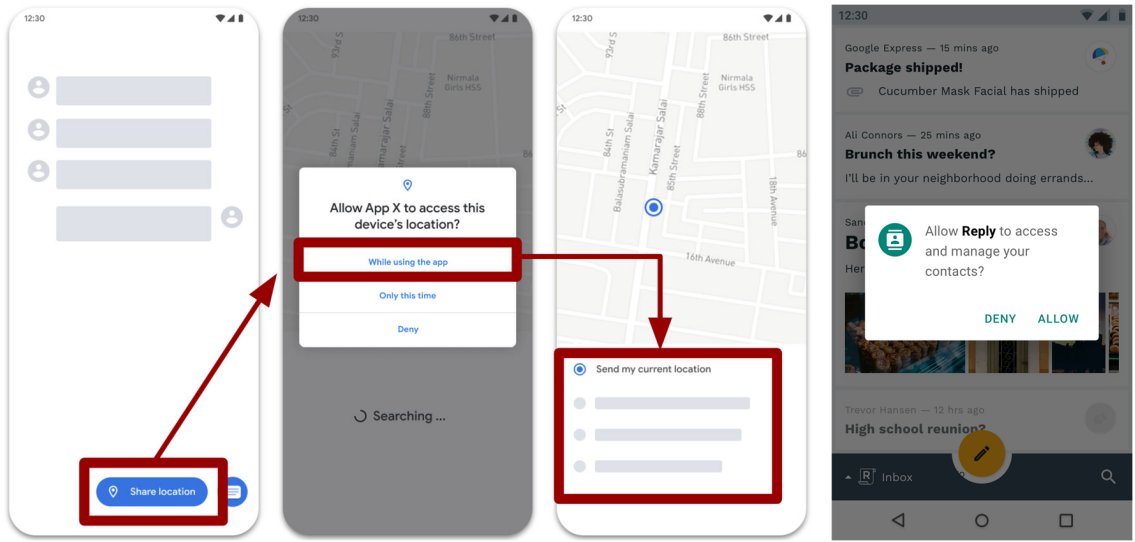

**Figure 11.** Android's current runtime permission user interfaces. The left comes from Android documentation (https://developer.android.com/training/location/permissions) and shows an example of how the runtime permissions work for location data, which has the choices of "While using the app", "Only this time", and "Deny". The right example comes from Material Design (https://material.io/design/platform-guidance/android-permissions.html) and shows how runtime permissions work for other permissions, which offer the choices of "Deny" and "Allow".

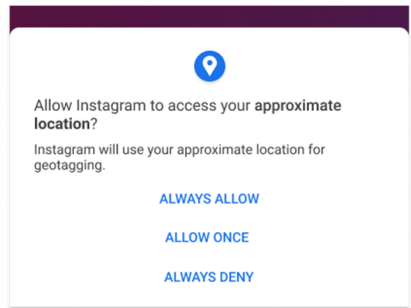

**Figure 12.** PE for Android's Runtime Permission user interface.

### 5.8 Runtime User Interface

Both iOS and Android currently support runtime user interfaces, where the operating system asks users about the use of sensitive information when the app needs it. Figure 14 shows two examples of Android's current user interfaces.

Figure 15 shows our design for Runtime permissions. This design is a variant of the policy card presented in Section 5.1. Our design shows the permission and purpose (both based on a fixed set of options).

Runtime UIs are a user setting, and thus are enforced with the lowest priority in CMU's policy enforcement algorithm. If an organizational profile or quick setting are in effect, or if there is a relevant App-specific Setting or Global Setting that allows or denies, a runtime prompt will not be shown.

As with the policy cards example shown in Figure 7, the purpose string from the app policy (see Figure 5 and Table 2) is also missing here, again due to time and prioritizing other parts of the design. Ideally, the purpose string would also be shown on this Runtime user interface to let developers give a short explanation of data use. Note that this might cause some issues with layout, however, and so there should be a maximum length for these strings.

We considered making this Runtime user interface look exactly like the policy cards shown earlier (see Figure 7), allowing people to move the slider to allow, ask, or deny and then proceed. However, informal feedback from user tests found that this approach felt complex, and so we went with a simpler design.

We considered including options for allowing background / foreground access to data, similar to Figure 13 left. We didn't include it in our design mostly for reasons of time and priorities, and because the existing Android design seems to do a good job already.

In short, the Runtime user interface could start with the existing Android design, adding purposes and the purpose string.

### 5.9 System Architecture for the Policy Manager

Here, we offer a high level view of the system architecture for the Policy Manager (see Table 3 and Figure 16).

The `DataRepository` centrally manages all data for the policy manager. There are two modes to the `DataRepository`: persistent and in-memory. Persistent is used for when the app is actually running (and is the default mode), and in-memory is used when testing the policy manager during development. It exposes an API to let outside components read/write/update to storage. The `DataRepository` also emulates data sources such as the Google Play store, which is needed to store off-device policies for apps we have analyzed.

The `PolicyManager` component exposes APIs for determining a privacy policy to enforce, as well as adding, fetching and updating those policies. The `PolicyManager` communicates via a `UserPolicy` data structure, which contains: app name, permission, purpose, third-party library or library category (if applicable) and policy action (allow, deny or ask). The `PolicyManager` component also manages the active policy profile, and installs them.

- The `PolicyManagerService` is an abstract PE for Android component, which every policy manager on a PE for Android device must implement. It exposes three communication points with PE for Android that our policy manager interacts with:
- `onAppInstall`: PE for Android will use this to tell us that an app has been installed, and gives us the off-device policy JSON string if the app is bundled with one.
- `onDangerousPermissionRequest`: when an app requests permission for privacy-sensitive data, PE for Android will forward that request to our policy manager by using `onDangerousPermissionRequest`, which has to determine if the access should be allowed or not. PE for Android gives us information such as the app requesting permission, the permission it is requesting, a purpose (if the developer provided one via the PE for Android SDK), stacktraces and a few other things. Policy enforcement algorithms are run here.
- `onPrivateDataRequest`: when a micro-pal module (see https://android-privacy.org/) requests sensitive data on behalf of an app, PE for Android will forward that request to our policy manager where it will determine if this access is allowed or not. Policy enforcement algorithms are run here.

| Component Name | Description |
|---|---|
| DataRepository | Centrally manages all storage and external data fetching for the policy manager. |
| PolicyManager | Exposes an API to manage policy profiles, and CRUD operations on UserPolicy objects. |
| PolicyManagerApplication | Manages and defines the current set of policy manager user interfaces. Also initializes the policy manager application as a whole. |
| PolicyManagerService | PE for Android component, which exposes three hooks: onAppInstall, onDangerousPermissionRequest and onPrivateDataRequest. Each hook is called via PE for Android, and delivers some response back to PE for Android. |

**Table 3.** Major components for the Policy Manager.

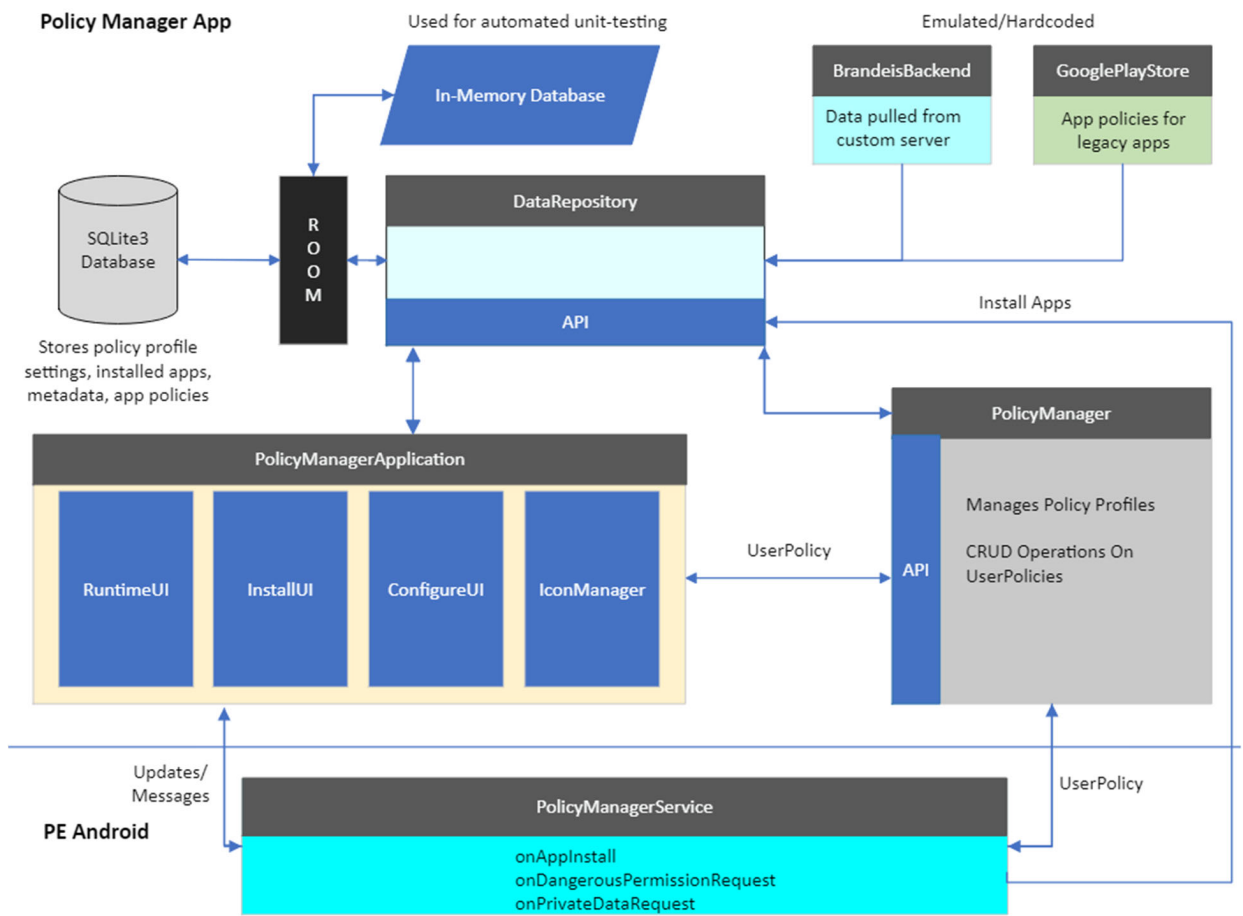

**Figure 13.** Architecture diagram for the Policy Manager.

During the course of enforcing privacy policies, `onDangerousPermissionRequest` or `onPrivateDataRequest` will communicate with the user interfaces to silently inform the user that a policy has just been enforced, or to prompt the user for a policy decision at runtime. The policy enforcement algorithms that run here will communicate with the `PolicyManager` component to determine active policy profiles, and which privacy policy action (allow or deny) that should be sent back to PE for Android. Lastly, the `PolicyManagerService` component will communicate with the `DataRepository` in order to determine things like installed apps or which off-device policy should be used if an app does not provide one.

The `PolicyManagerApplication` component abstracts the policy manager as a general application, which is broken into user interface modules. This allows for multiple implementations of the user interfaces, which the user can then switch between during runtime. The original intent of this component was to allow us to experiment with multiple implementations of policy managers on one device, but this was an avenue we ultimately did not pursue. This modularity also affords us the ability to compare different user interface designs/implementations when conducting user studies, without requiring users to return a device to install a new application.

The user interface modules are broken down into the major components a policy manager needs: runtime, install, configure and icon manager. Each interface reads, writes and updates data from the `PolicyManager` and `DataRepository` components so it can populate widgets and controls in their screens. The modules may also interact with the `PolicyManagerService` component when an app is being installed, a user needs to be prompted for a policy decision or notifications are sent out when a policy is enforced.

## 6 USER STUDIES

We conducted a series of user studies to inform the design of our user interfaces. This design process was carried out iteratively over time as insights and findings from user studies continued shaping and refining our designs. Initially, a few user studies explored user feedback for defining the information architecture and layout of the interfaces. We designed multiple preliminary UIs and then conducted more user studies that examined the understandability of policy card components and the relation between them. Feedback from the studies helped refine the designs further, and the later studies evaluated the effectiveness and usability of privacy manager for various key user flows. The following elaborates on the user studies and explains the design process.

In the initial phase, we focused on narrowing down to a structure/hierarchy that can represent the nested information of sensitive data access in an understandable and actionable format. A card sorting activity with participants showed that 80% of them preferred the what-why-who sequence of understanding the information, which informed the design of the policy card's structure. Following that, multiple rounds of A/B Testing of user interface prototypes gave rise to the zooming user interface. The zooming UI aimed at mitigating information overload for users by providing progressive disclosure of information on demand - i.e., giving them a short overview of what sensitive data the app will use and then the option to view more details and to configure permission settings on demand.

In the next phase, a preliminary version of the UI was designed and then evaluated by more user studies to gauge the understandability of the design. First, a comprehension survey was created and sent to 70+ MTurk users. Then a participatory design activity focused on digging deeper into the comprehension of the design was conducted with 7 in-lab users. These studies showed broad themes of difficulty in understanding the taxonomy and purpose of data access, discovering information, and differentiating between different types of policy levels.

More technically sounding permission names and purposes led to confusion and a lack of understanding of the data access among users. They wanted to see more descriptions of those terms for better clarity behind the data access. Next, no uniformity and consistency in the naming led to low differentiation between the policy levels, such as, most commonly seen, mistaking purposes for permissions. Hence it became necessary that the taxonomy and hierarchy should indicate the policy level that can help users understand the nested structure of information. Lastly, as the 'why' and 'who' are new types of policies being presented for configuring privacy, users didn't expect to find them, and they went unseen due to low discoverability in the design. Effective discoverability of policies would ensure an increase in user awareness of the levels of information they can view and control. Overall, we observed high learning curves for using the system, but most users appreciated having the availability of the granular control of privacy and found the policy cards useful.

Our team also conducted exploratory interviews to uncover existing pain points and needs that users face with current Android privacy controls so as to design features that will increase usefulness for users. The most common pain points reflected were a cumbersome experience to find the right settings, difficulty in remembering previous settings, no transparency in what is precisely being collected, and feeling distressed over giving out data to unknown parties. Through PE Android's capabilities, a lot of these problems could now be addressed, but it required that the design supports the user's ability to make decisions and take actions. This led us to improve the app settings and global settings' user flows to make the control understandable and quick to access while ensuring users feel confident of securing their data. These insights also helped us shape the recommendation feature and privacy overview graphs. Recommendations help by suggesting users to configure settings based on their past behavior and overview graphs help by giving users an overall impression of their data access and highlighting uncommon or unusual behavior.

The most complicated flows were app settings and global settings as they present the policy cards in great detail and require a thorough understanding on the user's end before they can make informed decisions. In the last phase, we developed two concepts to present information to users for both the flows and conducted concept testing with 10 participants to understand what design aspects help users achieve their privacy goals. For app settings, we wanted to learn if users care to control and organize all permissions of an app based on app internal use and third-party use or if they care to control these settings based on one permission at a time. Most users preferred the latter version of organizing and grouping settings based on permission as it matched with their mental model of configuring settings and also helped align with policy cards' hierarchy. Easier to prioritize and take actions, users found it more comprehensible and mangeable. For global settings, we wanted to learn if users care to control settings based on purposes across multiple apps collectively or control permissions based on apps requesting it. Based on current android UIs, most users associated more with focusing apps first to control settings. However, most users also identified with purpose based control over multiple apps when they cared more about purposes such as advertising,etc. Seeing a list of apps under purposes also helped them compare them and gave them a better understanding of the usage and evaluated the association of apps and purposes.

The concept testing gave very useful insights to finalize designs for the flows. In our final usability testing, we presented the entire flows to users and gave them key tasks that would prove the most usefulness of our design, if they completed in a reasonable time frame. While the results of the testing showed that all users appreciated the granularity and found the overall control and flows useful, the learning curve overall for using the features was still high for new users as these settings and details of control were never seen by them before.

In our final design iteration, we kept this mind and introduced helper text animations and onboarding screens to educate the user about the capabilities of PE Android and the wide range of control they have at their disposal.

## 7 RESULTS AND DISCUSSION

In this section, we describe some limitations we found in extending Android's permission system, limitations of our App Policies, how to improve generation of App Policies for existing apps, and our work with respect to Apple's privacy nutrition labels.

### 7.1 Limitations of Extending Android's Permission System

To recap, the two core ideas underlying our work are to require developers to declare the purpose of use of sensitive data, and to separate permissions and purposes into first-party use and third-party use. In this white paper, we presented an overview of how these two ideas can help improve privacy in many ways, for example helping developers when coding, helping with app analysis, and better user interfaces that can explain to end-users why an app is requesting sensitive data. However, our approach has several limitations. In this subsection, we focus on limitations due to our reliance on Android's permission system.

First, there are many kinds of sensitive data that do not have a corresponding permission, and our approach currently does not work for these cases. Some examples include:
- smartphone-related information such as the smartphone's MAC address
- seeing what other apps are installed on one's smartphone
- personal data such as one's health condition or one's personal finances

Similarly, our approach does not offer any solutions to personally identifiable information that an app might collect, for example:
- identifiers that are generated by an app
- information that a person enters into the app (e.g. one's email address or age)
- behavioral data such as what articles a person is reading in a news app or how long they play a game
- app-specific data, such as browser history for a web browser app

Lastly, our approach does not help with any app-specific logic or user interfaces regarding privacy, for example an app that would like to share analytics of how you use a given app with Yelp or a health app that wants your approval before sharing health information with your doctor.

Second, our approach also does not address tunnelling, where other devices might use the smartphone network capabilities to send data to the cloud. For example, in 2017, the New York Times reported how one manufacturer collected usage information on a Bluetooth-enabled sex toy via their app (de Freytas-Tamura, 2017). These kinds of situations are likely to grow as Internet of Things devices are increasingly deployed.

Third, our approach also does not directly address legal and compliance issues for privacy. For example, the Children's Online Privacy Protection Act (COPPA) imposes constraints on what kinds of data apps can collect from children. While there have been a few systems built to inspect apps and look for likely COPPA violations (Liu et al, 2016; Reyes et al, 2017), developers still have little guidance in the design and implementation phase of app development and analytics phase of app deployment. More generally, developers have very little guidance and support for the increasing number of laws and regulations surrounding privacy.

We believe that our work here can be adapted to help with some of these cases. For example, for the first issue of sensitive data not associated with a permission, we could easily look for use of specific APIs that indicate use of sensitive data and require an explicit privacy annotation. It is also possible to expand the concept of privacy annotations to all data that flows out of the smartphone to third parties, which would then facilitate auditing by app developers and by third parties (see Section 8.2 for a brief discussion of support for auditing). These kinds of annotations might also be adapted to address the second problem of tunneling too, where a device would need to communicate to the smartphone what data it is trying to share. However, note that end-to-end encryption does pose challenges here to auditing, and there is currently no good solution to this problem.

For the third problem, legal compliance, one could have checklists that help developers better understand rules and regulations around common kinds of data collection, as well as tools that help generate appropriate user interfaces and privacy policies that can pass muster with Google Play as well as existing laws. We briefly describe some ideas in this space in Section 8.2.

### 7.2 Limitations of Our App Policies

Our current design for app policies makes it so that if a sensitive permission is requested in a given method, there is exactly one associated purpose. For example, we can support a single method requesting both "location for social media" and "contact list for backup", but cannot support a single method requesting both "location of social media" and "location for navigation". In many apps, sensitive data is only used for one purpose, so this limitation should only pose a minor inconvenience for most developers. For apps that use the same data for multiple purposes, however, apps may need to be re-architected so that there is a clear flow from when the sensitive data is accessed to how it is used. That is, developers should not, for example, pre-fetch location data and then have different parts of their app use that data for different purposes. In some of our past work (Wang et al, 2015), we did find a few apps that used permissions indirectly in this manner, but do not believe that this kind of design is common.

A limitation of our current implementation is that we assume the app policies are correct. For app policies that are generated for existing apps, this should not be a problem, since we assume that these app policies will be created by an app store or some other trusted party. The issue then is for app policies for new PE for Android apps. There are at least four points where app policies can be checked:

(1) when the app is initially compiled
(2) when the app is uploaded to an app store
(3) when the app is installed on a smartphone
(4) when the app is run on a smartphone

For scale and performance reasons, (1) and (2) are likely the best options. Option (2) in particular is intriguing since app stores are centralized, offering a single point where the privacy and security of the vast majority of apps can be checked (Gilbert et al, 2011). In Section 7.3, we sketch out a number of existing techniques that could be used to improve the generation of app policies for existing apps, which could easily be adapted to check the correctness of app policies embedded in new apps. These techniques could also be complemented with analyses that compare apps in aggregate, looking for patterns in how sensitive data is commonly used by apps. For example, Frank et al applied clustering techniques to permission requests, and found that low-reputation apps often had more complex permission requests that differed from high-reputation apps (Frank et al, 2012). Lastly, in Section 8.2, we describe some research ideas for improving the ability for developers, app stores, and third parties to audit apps, which could also help with checking the correctness of app policies.

Another limitation of our app policies is that they are focused on data being shared with cloud servers, as opposed to data being shared with other people (e.g. what a person might choose to share or accidentally share with others on social media). It's not clear if there is a good angle of attack here at the app policy level.

### 7.3 Improving Generation of App Policies for Existing Android Apps

Currently, we only use relatively simple static analysis techniques to generate app policies, inspecting what third-party libraries apps use, looking for the use of specific Android APIs and the package names of classes to infer likely purposes. In this section, we sketch out some ideas for improving the generation of app policies. In many respects, generation of app policies for existing Android apps is loosely coupled with the rest of our approach, meaning that if new techniques for app analysis are developed, they (and any new generated app policies) can be independently deployed on app stores without having to update PE for Android or its installed base. We also note that the same techniques to generate app policies for existing apps can also be used to check app policies for PE for Android apps, to make sure that developers are being honest and correct.

Generally, static analysis should be preferred over dynamic analysis due to reasons of scale. Given that there are several million Android apps in existence, it is infeasible to apply dynamic analysis techniques to all of them. Furthermore, past work has found major challenges with running apps using scripts or other kinds of UI monkeys (Amini et al, 2012; Lee et al, 2013; Amini 2014; Hao et al, 2014; Li et al, 2017), e.g. non-deterministic behavior such as ads or requests to rate the app, dynamically loaded content which differs each time such as news articles, infinite scroll, text fields that require the user to enter in something, creating and logging into user accounts, apps that make extensive use of Canvas rather than standard GUI widgets, as well as state explosion in general.

We believe a hybrid approach may be effective in generating app policies. For example, static analysis can be applied for all apps, with dynamic analysis being applied using automated UI monkeys to the top 10000 most popular apps, supplemented with manual interaction and manual inspection for the top 1000 most popular apps or ones that raise too many red flags. Given that there is likely a long tail of app installations, this hybrid approach would offer breadth as well as depth for analysis while also being judicious with analysis costs (in terms of server costs as well as human labor costs).

For static analysis, there are two major inputs. The first is the app metadata, such as an app's text description (Pandita et al, 2013; Gorla et al, 2014; Qu et al, 2014), privacy policy, app category, etc. The second is the app binary. As mentioned earlier, in our past work (Chitkara et al, 2017) we found that about 40% of apps only use sensitive data because of third-party libraries. This means that analyzing what libraries an app uses can go a long way for many apps. However, this also means that we need more techniques to analyze custom code. Some possibilities include decompiling an app and extracting the strings of class names and method names (Wang et al, 2015), as well as strings of URLs and website domains mentioned in the app. Flow analysis, using tools such as FlowDroid (Arzt et al, 2014), IccTA (Li et al, 2015), and FlowCog (Pan et al, 2018), can also be applied to understand information flow in apps.

One might also analyze code for potential red flags for privacy and then subject those apps for further manual inspection. Some examples include dynamic loading of classes, use of reflection to run code, native code called via JNI, as well as apps that have characteristics that do not conform to how "normal" a behavior is for that app category (Frank et al, 2012) or differ greatly from apps with a similar text description. For example, many navigation apps use GPS, but very few productivity apps do so.

Note that much of this static analysis assumes that the APK is not obfuscated. In practice, we have found that the vast majority of third-party libraries are not obfuscated (Chitkara et al, 2017). With respect to custom code, past work has found that only about 2% of apps have obfuscated code (Linares-Vásquez et al, 2014), and in our work on text mining class names and method names (Wang et al, 2015), we found that only 10% of apps had any obfuscated code.

With respect to dynamic analysis, there has been a great deal of past research in the context of smartphone apps and often with respect to malicious apps (Burguera et al, 2011; Wei et al, 2012; Zheng et al, 2012; Rastogi et al, 2013; Avdiienko et al, 2015; Xia et al, 2015; Yang et al, 2015; Jin et al, 2018; Ren et al, 2018). For the purposes of generating and checking app policies, the most important things to model are what data is being accessed, where it is going, and why. Understanding what data and where it is going is challenging but can be accomplished using taint tracking techniques, information flow, and/or monitoring network traffic for data egress, this last option of which also assumes that a man-in-the-middle can be inserted. A complementary approach is to trace the flow of sensitive data and examine the stack traces from when sensitive data is accessed to when it egresses to network, using the names in those stack traces to infer purpose of use. This is a variant of the static analysis text mining of class names and method names but adapted to runtime. In our team's past work, we have done some initial work in this area, focusing on the point of access (Wang et al, 2017) and for third-party libraries (Chitkara et al, 2017), but not tracing the full flow of information. Also note that this approach assumes that developers do not deliberately try to deceive this kind of text mining by using misleading class and method names.

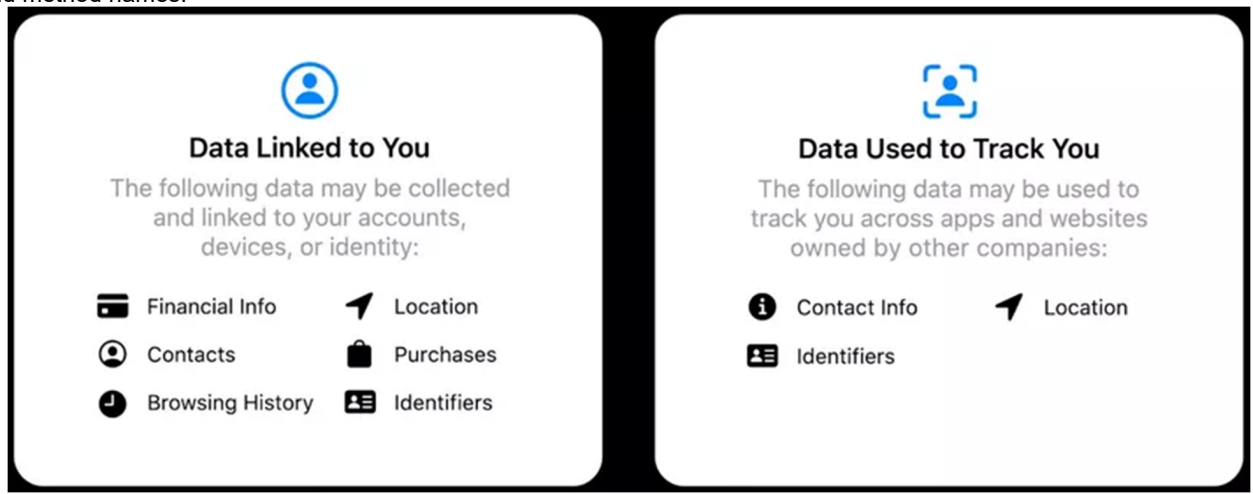

**Figure 14.** An example of Apple's privacy nutrition labels. These nutrition labels can be found on Apple's app store, though they currently require users to actively seek them out, and there does not currently seem to be any verification of the accuracy of the data.

### 7.4 Apple's Privacy Nutrition Labels

In late 2020, Apple started requiring that new and updated apps have a privacy nutrition label (see Figure 17). Developers must fill out information about what data is collected by an app, where collect is defined as data being transmitted off device. Developers must also disclose whether that data is linked to an individual and whether that data is used to track.

We believe many ideas from this paper could be used to help augment these privacy nutrition labels. One is with improving the accuracy of these labels. For example, one journalist examined several nutrition labels and compared them against network data and found several inaccuracies (Fowler, 2021). Privacy nutrition labels are currently manually filled out by developers, and are not directly connected to the source code of an app. We believe tightly coupling the source code to privacy summaries of an app's behaviors as well as privacy user interfaces may be a more effective approach, as described in Section 3.2 and expanded on in Section 8.2.

Another is with expanding the taxonomy of uses. Currently, Apple only has six kinds of data uses (Apple, n.d.), including Third-party advertising, Developer's advertising or marketing, Analytics, Product personalization, App functionality, and Other purposes. Our full taxonomy of purposes (see Appendix B) can offer more insight into how data is being used. Apple's taxonomy is heavily focused on tracking and advertising, which misses other kinds of privacy concerns, e.g. deployed military personnel accidentally sharing location data.

## 8 FUTURE WORK IN IMPROVING SMARTPHONE PRIVACY

In this white paper, we have presented our team's work on PE for Android's app policies and user interfaces. However, our work only represents one point in the entire space of possibilities for improving privacy. In this section, we sketch out several ideas for improving smartphone privacy more broadly, focusing on issues related to technology and the user experience rather than regulatory or policy approaches.

### 8.1 Reducing the Cost for End-Users of Managing Their Privacy

A major challenge for privacy today is that the burden of privacy is shouldered almost entirely by end-users. Individuals need to know a great deal about the capabilities and configurations of their devices, their operating systems, and the apps they use. Individuals also must have a great deal of awareness and knowledge as to how to protect themselves, and they must take many affirmative steps to enable such protection.

Breaking things down, some of the labor costs in managing one's own privacy include the attention costs to monitor what is happening with one's data (e.g. notifications) or to maintain how one is presenting oneself to others, the cognitive costs of understanding different privacy user interfaces (e.g. understanding a specific privacy issue, knowing that a specific privacy setting exists, or navigating an app's specific interfaces for privacy), the search costs to find the right user interfaces to manage one's privacy (e.g. finding the right setting), and evaluation costs of checking that a given setting was changed correctly and has the desired effect.

One approach for addressing these user costs is to have *simpler, coarse-grained approaches for privacy*. While fine-grained approaches (for example, being able to configure lots of different settings) offer a great deal of flexibility, they also incur a high cost for end-users in terms of understanding one's options and setting things correctly. Lederer et al (Lederer et al, 2004) instead ask developers to also consider offering coarse-grained controls. For instance, past work has investigated modifications to smartphone operating systems that return fake data when apps request sensitive data, such as a fake contact list or fake location (Beresford et al, 2011; Agarwal and Hall, 2013). One could imagine offering this functionality for smartphones as an incognito mode similar to that offered by web browsers, making it easy for users to try out apps without fear of leaking data, or just temporarily hiding one's personal data, without having to adjust multiple knobs and dials. The challenge here is for developers to determine what a good set of settings are in a way that will not break most apps, and for designers to convey these semantics to end-users.

Another approach for addressing these user costs is to *use crowdsourcing to assess people's mental models and use aggregated privacy decisions for recommendations*. The crowd here might be actual users of an app or workers being paid to examine some aspect of an app. For example, people might be asked to answer multiple choice questions about the behavior of an app before installing it (e.g. does this app use your location data for advertising?), and be shown the answer if incorrect. If many people answer the same question incorrectly, this suggests that there is a potential major mismatch between people's expectations of an app's behavior and reality. In past work, we used a variant of this idea, asking crowd workers via Mechanical Turk about their expectations (Lin et al, 2012; Lin et al, 2014). Another approach is to collect people's decisions as to whether or not to share requested data, and then show the results to other users in the future (Agarwal and Hall, 2013). For example, "85% of people chose to share microphone data with this app". It is also possible to cluster the crowd's decisions into different groups and then determine which cluster is the best match for a person (Lin et al, 2014).

### 8.2 Better Programming Models and Developer Tools for Privacy

A complementary approach to helping end-users with privacy is to help developers with privacy. For example, in our team's past research, we conducted surveys and interviews with developers and found that the vast majority have very little knowledge of best practices and regulations around privacy (Balebako et al, 2014; Li et al, 2018). However, while there is a great deal of past work in terms of analyzing smartphone apps with respect to privacy and security, there is very little support in terms of helping developers with little knowledge of these do better.

One idea here is to *expand the functionality and utility of the privacy annotations* previously described in Section 3.2. The broad strategy here is that if developers can do a little bit of extra work with these annotations, it can not only make the developers' lives easier, it can also vastly improve the privacy ecosystem in a number of ways. For example, these annotations might be used to automatically generate privacy user interfaces or human-readable privacy policies that can pass muster with Google Play. Another idea is to expand on the auditing functionality in Coconut, making it much easier for a development team to see all of the conditions in which sensitive data might be collected and where that data goes (discussed more below). Lastly, these privacy annotations might also be directly embedded into compiled code to help with app analysis, auditing, and enforcement by others, e.g. journalists, researchers, privacy advocates, and app stores. These are all ideas we are exploring in our team's ongoing research.

A complementary idea here is to *offer better programming models for privacy*. For instance, in our past work on PrivacyStreams (Li et al, 2017), we offered a stream-based API for accessing sensitive data on Android. One key insight is that there is a mismatch between the data that Android offers via its APIs and the granularity of data that developers need. For example, we found that many apps request GPS data, but actually need neighborhood or city granularity. As another example, an app might request access to all SMS messages, but only need messages from a single sender for two-factor authentication. Another key insight is that, by offering a stream-based API, we can make it easier for developers to get the granularity they want while also improving privacy almost as a side effect.

For example, below shows an example of how a developer can request access to microphone, set up a series of transformations to get loudness, and then send the result to a callback.

```
uqi.getData(Audio.recordPeriodic(10*1000, 2*60*1000),
            Purpose.HEALTH("sleep monitor"))
   .setField("loudness", AudioOperators.calcLoudness("audio_data"))
   .onChange("loudness", callback)
```

Here, the developer does not have to deal with threading and buffers, only needs to learn a single uniform API to access any kind of sensitive data, and can easily get the granularity they want by specifying a chain of operators to filter, aggregate, or transform the data. At the same time, this stream is easy to statically analyze, making it possible for us to generate human-understandable text like "this app uses microphone to get loudness". It is also possible to get stronger guarantees, in that the sensitive data can be processed in an isolated query engine or a separate process space with only the final result being returned to the app. See https://privacystreams.github.io for more details.

We have a few comments here about PrivacyStreams. First, a variant of PrivacyStreams is incorporated into PE for Android, though it does not support chained operators. Second, there are several challenges to making a query engine for sensitive data work for Android. For example, let's assume the query engine is an always-on background service. While it makes sense for the query engine to be able to access any sensitive data on one's smartphone, it also needs to be limited by what permissions the app making the query has explicitly declared in its manifest. Otherwise, an app could bypass the permission system and access any data.

Another major opportunity for helping developers is *better support for auditing the behavior of their apps*. In some of our interviews, we found that some developers did not know the full extent of data collected by their apps (Agarwal and Hall, 2013; Li et al, 2018). Some of the explanations include third-party libraries with unclear behaviors, different versions of apps collecting different kinds of data, large distributed teams with people working on different parts of the app, and turnover in those teams leading to a loss of knowledge. However, even if developers did know the full extent of the data collection and data uses of their apps, auditing needs to go beyond just the immediate development team. For example, we need to be able to support Chief Privacy Officers, Quality Assurance, and people in Legal roles to check that apps are complying with all laws and regulations as well as with one's own stated privacy policies. One such example would be automatically generating tables of all of the data that an app collects, the associated conditions (e.g. relevant screenshots), the names of the variables that data is sent as via network traffic (i.e. the keys in key-value pairs in the message payload), where that data is sent to (e.g. our servers, other cloud service providers), and so on.

Lastly, our work on app policies is just one point in the design space. Having *more kinds of high-level constraints for apps* can greatly facilitate system design, enforcement, and analysis. For instance, IETF RFC8520 proposes Manufacturer Usage Descriptions (MUDs) for Internet of Things devices, which are essentially whitelists of network domains that devices will communicate with. The rationale here is that the vast majority of devices only need to contact a few sites, for example for software updates or to offer services, and so declaring a whitelist of sites can greatly improve security. This idea of whitelists also makes a great deal of sense for apps. In fact, this idea could be linked with the above ideas of privacy annotations and support for auditing, where an app must declare what remote servers it will communicate with, as well as the full range of data that will be sent to each of those servers. Other possibilities for high-level constraints include specifying whether the data will be accessed by the app in the foreground or background, whether the data will be accessed automatically or due to user interaction such as clicking on a button (Huang et al, 2014; Li et al, 2016), and how frequently the data will be accessed (e.g. only once, only on load, continuously, etc). This kind of explicitly declared metadata would make it much easier to analyze an app's behavior and also create new opportunities for explaining to end-users what an app will and won't do.

### 8.3 Improving the Motivation of Developers and Companies for Privacy

Das et al introduce the concept of security sensitivity for end-users, describing how end-users need a combination of awareness, knowledge, and motivation to successfully adopt new security tools and practices (Das et al, 2014). These same concepts can be adapted to describe what developers need to successfully adopt new privacy tools and practices. The previous subsection described how developers have low awareness and knowledge of privacy, and how new kinds of tools and programming models might address this problem. Here, we look at the issue of motivation.

Currently, privacy has clear costs on the part of developers and companies in terms of time and effort to understand what's required, learning new tools and changes to APIs, designing appropriate architectures and user interfaces, implementation, testing to make sure the app still works as intended, and auditing to check what sensitive data is accessed and how it is used. However, there are unclear rewards and benefits. In some ways, privacy is a market failure. For example, a person might purchase a web camera based on easily discernible features like color, size, and price, but privacy rarely enters the picture. This is due in part to the fact that privacy is intangible, hard to assess, and hard to understand. Another way of describing the current landscape is that, for app developers and companies, there are only sticks and no carrots.

The research community has much to offer here, in terms of investigating the effectiveness of different rewards for privacy. The most obvious point of leverage is the app store itself. For example, when searching for apps, the ranking algorithm of the app store could take relatively simple privacy-related metrics into account, such as the amount of data collected and the number of invasive third-party libraries used. App stores could also make privacy information more prominent on individual app pages, and make it easier to compare apps with similar functionality based on factors like number of downloads, star ratings, and privacy. In all of these cases, researchers could investigate what metrics people find most useful, and how those metrics influence people's decisions to download and use apps.

Researchers could also investigate how privacy might mesh with branding, or the effectiveness of privacy as a core value proposition compared with competitors. Similarly, researchers could also examine if certain kinds of user interfaces or functionality get people to talk more positively about privacy, to help with word of mouth or social media. For example, past work has found that people rarely talk about privacy in app store reviews, and usually the feedback is negative (Ha and Wagner, 2013; Fu et al, 2013). Are there ways to increase discussion of privacy in app store reviews or on social media, and for people to offer positive feedback instead of just negative feedback? Also, how do these discussions impact likelihood to download and use, as well as overall user experience?

### 8.4 A Community Resource for Analyzing Apps and Sharing Analysis Code and Data

Our final idea here is to build out a community resource that would make it easy for anyone in the world to analyze Android apps and share their code and results. Researchers in a wide range of fields are analyzing smartphone apps, examining issues of privacy (Lin et al, 2014; Lin et al, 2014; Ren et al, 2018; van Kleek et al, 2018), security (Frank et al, 2012; Zhou et al, 2012; Egele et al, 2013; Rastogi et al, 2013; Fischer et al, 2017; Jia et al, 2017), accessibility (Ross et al, 2018; Alshayban et al, 2020), software engineering (Viennot et al, 2014; Wang et al, 2017), in-car safety (Lee et al, 2013), and more. However, many of these analyses only examine a few hundred or a few thousand out of the several million apps that are available, in part due to the challenges of downloading these apps as well as setting up the tools and infrastructure for analyzing them. Furthermore, relatively little software or raw data has been shared, making it slow and difficult for researchers to build on top of each others' work.

We argue that a centralized community hub could greatly streamline analysis and sharing of results, and thus improve the privacy and security of the entire app ecosystem. This hub would crawl apps from Google Play and keep those mostly up to date, and make it easy to do certain kinds of static and dynamic analysis. Researchers around the world could upload their analysis code to this hub and run it on a sample or all of the apps (subject to some security constraints to block malicious attackers). When those researchers are ready, they could flip a switch and easily share their code and results with everyone in the world. This would make it easier for people to replicate studies, rigorously compare analysis techniques, and build on past results. Such a community resource would also greatly lower the barriers for government agencies, journalists, consumer advocates, and privacy advocates to analyze and audit apps.

## 9 CONCLUSION

In this white paper, we presented the user interface design and rationale behind new user interfaces for Privacy Enhancements for Android, a DARPA-supported project. We sketched out how our two core ideas of having developers explicitly declare the purpose of use and of separating between first- and third-party use of data can improve many aspects of the entire smartphone privacy ecosystem.

We also sketched out several directions for future research, including ways of reducing the burden of privacy on end-users, helping developers do better with respect to privacy, and building out a community resource to facilitate app analysis and sharing of results.

We are making all of our designs and our source code public on our web site at https://android-privacy-interfaces.github.io in the hopes of improving the privacy landscape for smartphones, and believe that many of our ideas are applicable to emerging technologies that are likely to have similar privacy concerns, for example Augmented Reality and Internet of Things.

### ACKNOWLEDGMENTS

This material is based on research sponsored by Air Force Research Laboratory under agreement number FA8750-15-2-0281. The U.S. Government is authorized to reproduce and distribute reprints for Governmental purposes notwithstanding any copyright notation thereon. The views and conclusions contained herein are those of the authors and should not be interpreted as necessarily representing the official policies or endorsements, either expressed or implied, of Air Force Research Laboratory or the U.S. Government.

Elements of this research was also supported by Google through Google Faculty Research Awards. Special thanks to the many undergrads, masters students, and visiting scholars who participated in many of our discussions, including Shahriyar Amini, Gokhan Bal, Rebecca Balebako, Fanglin Chen, Saksham Chitkara, Kevan Dodhia, Nishad Gothoskar, Yao Guo, Suhas Harish, Yuan Cindy Jiang, Kevin Ku, Evelyn Kuo, Su Mon Kywe, Yuanchun Li, Jialiu Lin, Minxing Liu, Minyi Liu, Yile Clara Liu, Yejun Amy Lu, Asit Parida, Haley Park, Gaurav Srivastava, Weijia Rosie Sun, Haoyu Wang, Alex Yu, Tiffany Yu, and Max Zhu. Also, special thanks to members of the staff at Google and Apple who have given us feedback over the years. Lastly, thank you to the members of the DARPA Brandeis Mobile Collaborative Research Team for many insightful discussions and incredibly useful feedback over the course of this research.

**APPENDIX A    LIST OF ANDROID DANGEROUS PERMISSIONS**

This table lists dangerous permissions for Android, ones that are specially handled by PE for Android and checked against the Policy Manager. This list was taken from Android documentation. However, this list of dangerous permissions does not seem to be publicly available anymore for reasons that are unclear to us.

| | |
|---|---|
| CALENDAR | READ_CALENDAR<br>WRITE_CALENDAR |
| CAMERA | CAMERA |
| CONTACTS | READ_CONTACTS<br>WRITE_CONTACTS<br>GET_ACCOUNTS |
| LOCATION | ACCESS_FINE_LOCATION<br>ACCESS_COARSE_LOCATION |
| MICROPHONE | RECORD_AUDIO |
| PHONE | READ_PHONE_STATE<br>READ_PHONE_NUMBERS<br>CALL_PHONE<br>ANSWER_PHONE_CALLS<br>READ_CALL_LOG<br>WRITE_CALL_LOG<br>ADD_VOICEMAIL<br>USE_SIP<br>PROCESS_OUTGOING_CALLS |
| SENSORS | BODY_SENSORS |
| SMS | SEND_SMS<br>RECEIVE_SMS<br>READ_SMS<br>RECEIVE_WAP_PUSH<br>RECEIVE_MMS |
| STORAGE | READ_EXTERNAL_STORAGE<br>WRITE_EXTERNAL_STORAGE |

**APPENDIX B - TAXONOMY OF PURPOSES**

This table lists all of the purposes that our team has enumerated. Each purpose name aims to have a clear action inside of it, and has the word "For" prepended to it when displayed to make clearer that it is a purpose (e.g. "For Displaying Advertisements").

| Purpose | Description | Likely Permission(s) Used |
|---|---|---|
| For Displaying Advertisements | To deliver ads targeted to your interests. Examples include AdMob and Millennial Media. | `ACCESS_FINE_LOCATION`<br>`ACCESS_COARSE_LOCATION`<br>`READ_CONTACTS`<br>`RECORD_AUDIO` |
| For Gathering Analytics | To capture and analyze the behavior of apps and/or users, e.g. Flurry and Crashlytics. | `ACCESS_FINE_LOCATION`<br>`ACCESS_COARSE_LOCATION` |
| For Monitoring Health | To track and analyze your health habits | `ACCESS_FINE_LOCATION`<br>`ACCESS_COARSE_LOCATION`<br>`READ_CONTACTS`<br>`BODY_SENSORS`<br>`RECORD_AUDIO`<br>`READ_CALENDAR`<br>`WRITE_CALENDAR` |
| For Connecting with Other People or Social Media | To connect with or find other users, or functionality related to social media, e.g. Twitter and Facebook | `READ_CONTACTS`<br>`ACCESS_FINE_LOCATION`<br>`ACCESS_COARSE_LOCATION`<br>`CAMERA`<br>`RECORD_AUDIO`<br>`READ_CALENDAR`<br>`WRITE_CALENDAR`<br>`READ_CALL_LOG`<br>`ANSWER_PHONE_CALLS`<br>`CALL_PHONE`<br>`SEND_SMS`<br>`RECEIVE_SMS`<br>`READ_SMS` |
| For Conducting Research | To gather data for research studies or experiments | `Any permission` |
| For Backing-up to Cloud Service | To back up or save important data to a remote service | `READ_SMS`<br>`READ_CONTACTS`<br>`READ_EXTERNAL_STORAGE` |
| For Navigating to a Destination | To provide directions or guidance on how to travel somewhere | `ACCESS_FINE_LOCATION`<br>`ACCESS_COARSE_LOCATION` |
| For Searching Nearby Places | To search for businesses or events near you | `ACCESS_FINE_LOCATION`<br>`ACCESS_COARSE_LOCATION` |
| For Delivering Local Weather | To give weather reports for your immediate area | `ACCESS_FINE_LOCATION`<br>`ACCESS_COARSE_LOCATION` |
| For Adding Location to Photo | To tag photos with your location (geotagging) | `ACCESS_FINE_LOCATION`<br>`ACCESS_COARSE_LOCATION`<br>`CAMERA` |
| For Playing Games | To implement in-game features<br>(Note that this does not encompass game engines, which is under Development.) | `ACCESS_FINE_LOCATION`<br>`ACCESS_COARSE_LOCATION`<br>`RECORD_AUDIO`<br>`CAMERA`<br>`READ_CONTACTS`<br>`READ_CALL_LOG`<br>`READ_PHONE_STATE`<br>`READ_EXTERNAL_STORAGE`<br>`WRITE_EXTERNAL_STORAGE` |
| For Securing Device | To secure your device from unwanted persons | `READ_SMS`<br>`READ_CONTACTS`<br>`CAMERA`<br>`RECORD_AUDIO` |
| For Messaging or Calling People | To communicate with other individuals | `READ_SMS`<br>`SEND_SMS`<br>`RECEIVE_SMS`<br>`RECORD_AUDIO`<br>`ADD_VOICEMAIL`<br>`CALL_PHONE`<br>`READ_PHONE_STATE`<br>`READ_CALL_LOG`<br>`PROCESS_OUTGOING_CALLS` |
| For Recognizing Voice or Speech | To detect a speaker or audio commands | `RECORD_AUDIO` |
| For Streaming Media | To stream live audio, video, or both | `RECORD_AUDIO`<br>`CAMERA` |
| For Notifying Emergency Services | To contact emergency services | `ACCESS_FINE_LOCATION`<br>`ACCESS_COURSE_LOCATION`<br>`SEND_SMS`<br>`CALL_PHONE` |

| For Supporting Development | For helping developers build apps (Examples include graphics or audio processing) | `RECORD_AUDIO`<br>`CAMERA`<br>`ACCESS_FINE_LOCATION`<br>`READ_EXTERNAL_STORAGE`<br>`WRITE_EXTERNAL_STORAGE` |
|---|---|---|
| For Running Other Features | To run some basic or undetermined app feature. This is the default purpose if no purpose is specified or if we cannot determine the purpose. Should be used sparingly. Can also be used by static and dynamic analysis as potential red flag) | `Any permission` |

## APPENDIX C - EXAMPLE APP POLICY

Below is a longer example app policy for WhatsApp. Again, this app policy is expected to be embedded inside of an APK.

```
[
{
        "method": "A0q",
        "uses": "android.permission.GET_ACCOUNTS",
        "purpose": "Backup to Cloud Service",
        "for": "To back up or save important data to a remote service.",
        "class": "com.whatsapp.gdrive.RestoreFromBackupActivity"
},
{
        "method": "AD4",
        "uses": "android.permission.GET_ACCOUNTS",
        "purpose": "Backup to Cloud Service",
        "for": "To back up or save important data to a remote service.",
        "class": "com.whatsapp.gdrive.RestoreFromBackupActivity"
},
{
        "method": "A0X",
        "uses": "android.permission.GET_ACCOUNTS",
        "purpose": "Backup to Cloud Service",
        "for": "To back up or save important data to a remote service.",
        "class": "com.whatsapp.gdrive.SettingsGoogleDrive"
},
{
        "method": "onCreate",
        "uses": "android.permission.GET_ACCOUNTS",
        "purpose": "Running Other Features",
        "for": "To run some basic or undetermined app feature.",
        "class": "com.whatsapp.registration.RegisterName"
},
{
        "method": "A0l",
        "uses": "android.permission.SEND_SMS",
        "purpose": "Running Other Features",
        "for": "To run some basic or undetermined app feature.",
        "class": "com.whatsapp.payments.ui.IndiaUpiDeviceBindActivity"
},
{
        "method": "onCreate",
        "uses": "android.permission.GET_ACCOUNTS",
        "purpose": "Running Other Features",
        "for": "To run some basic or undetermined app feature.",
        "class": "com.whatsapp.accountsync.LoginActivity"
},
```

```
{
        "method": "*",
        "uses": "android.permission.READ_PHONE_STATE",
        "purpose": "Running Other Features",
        "for": "To run some basic or undetermined app feature.",
        "class": "*"
},
{
        "method": "*",
        "uses": "android.permission.RECEIVE_SMS",
        "purpose": "Running Other Features",
        "for": "To run some basic or undetermined app feature.",
        "class": "*"
},
{
        "method": "*",
        "uses": "android.permission.CAMERA",
        "purpose": "Running Other Features",
        "for": "To run some basic or undetermined app feature.",
        "class": "*"
},
{
        "method": "*",
        "uses": "android.permission.ACCESS_COARSE_LOCATION",
        "purpose": "Running Other Features",
        "for": "To run some basic or undetermined app feature.",
        "class": "*"
},
{
        "method": "*",
        "uses": "android.permission.ACCESS_FINE_LOCATION",
        "purpose": "Running Other Features",
        "for": "To run some basic or undetermined app feature.",
        "class": "*"
},
{
        "method": "*",
        "uses": "android.permission.READ_CONTACTS",
        "purpose": "Running Other Features",
        "for": "To run some basic or undetermined app feature.",
        "class": "*"
},
{
        "method": "*",
        "uses": "android.permission.RECORD_AUDIO",
        "purpose": "Running Other Features",
        "for": "To run some basic or undetermined app feature.",
```

```
            "class": "*"
    },
    {
            "method": "*",
            "uses": "android.permission.WRITE_CONTACTS",
            "purpose": "Running Other Features",
            "for": "To run some basic or undetermined app feature.",
            "class": "*"
    },
    {
            "method": "*",
            "uses": "android.permission.WRITE_EXTERNAL_STORAGE",
            "purpose": "Running Other Features",
            "for": "To run some basic or undetermined app feature.",
            "class": "*"
    },
    {
            "method": "*",
            "uses": "android.permission.CALL_PHONE",
            "purpose": "Running Other Features",
            "for": "To run some basic or undetermined app feature.",
            "class": "*"
    },
    {
            "method": "*",
            "uses": "android.permission.ANSWER_PHONE_CALLS",
            "purpose": "Running Other Features",
            "for": "To run some basic or undetermined app feature.",
            "class": "*"
    },
    {
            "method": "*",
            "uses": "android.permission.READ_CALL_LOG",
            "purpose": "Running Other Features",
            "for": "To run some basic or undetermined app feature.",
            "class": "*"
    },
    {
            "method": "*",
            "uses": "android.permission.READ_EXTERNAL_STORAGE",
            "purpose": "Running Other Features",
            "for": "To run some basic or undetermined app feature.",
            "class": "*"
    }]
```

**APPENDIX D - FINAL UI DESIGNS**

All of our user interface designs and code is available on our project website https://android-privacy-interfaces.github.io. Our user interfaces are also available on Figma at
https://www.figma.com/file/TDJXudKC9cjWH2Ex4M2c6BAH/Brandeis-19Spring?node-id=4034%3A14601

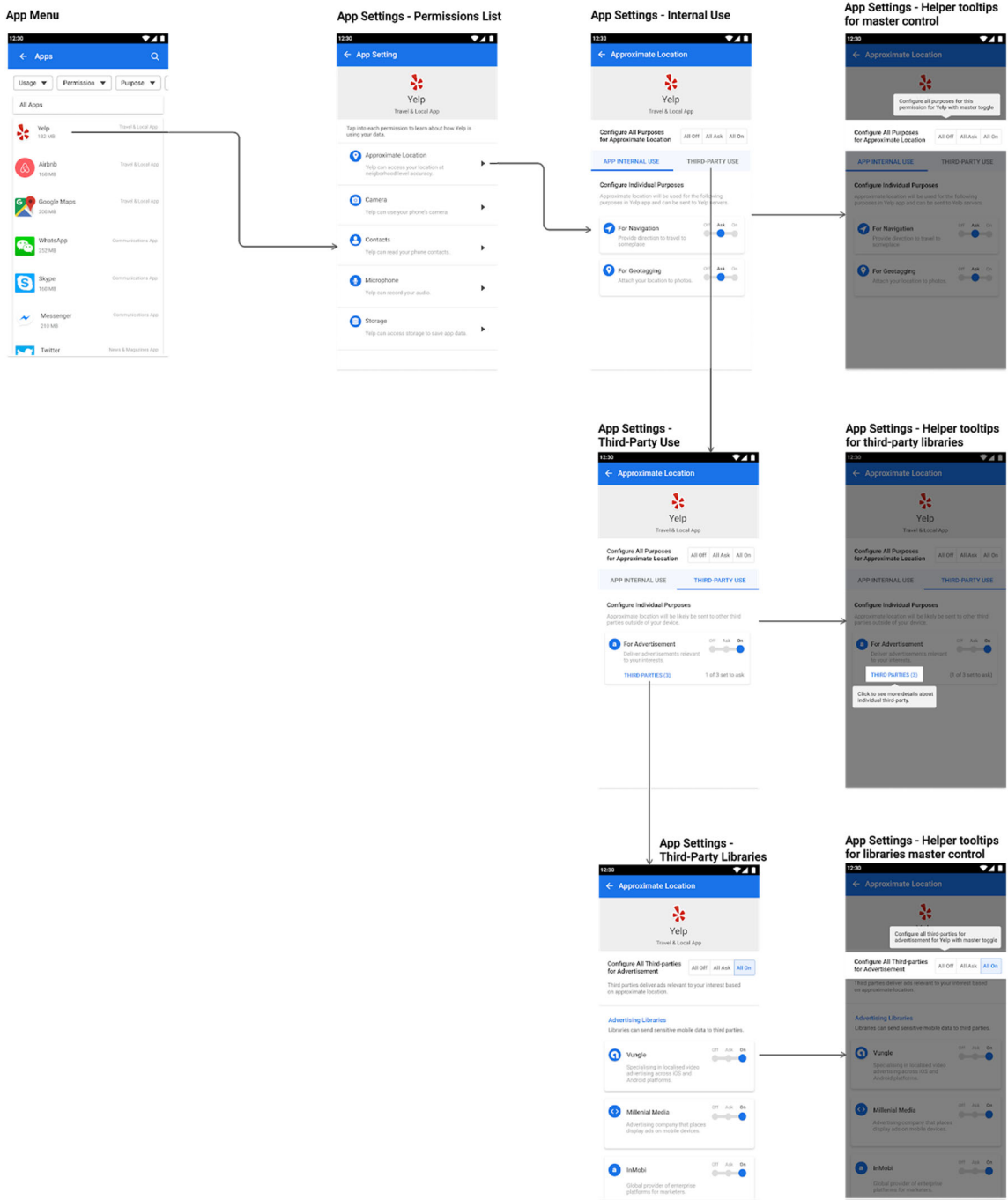

**Figure 15.** These screens show the interaction flow for App-Specific Settings.

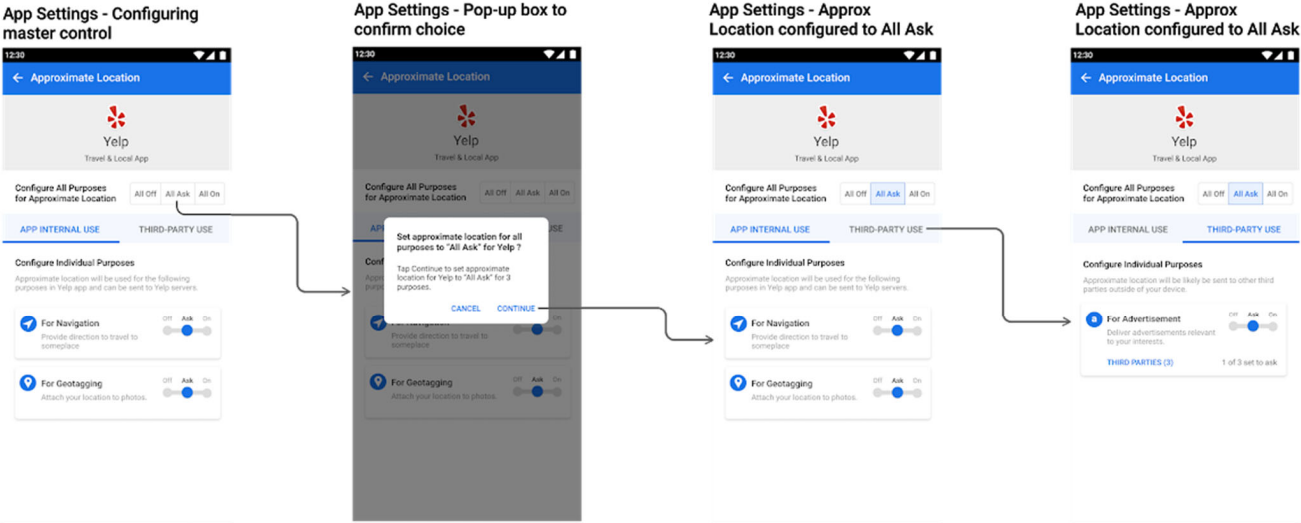

**Figure 16.** These screens show the interaction flow for setting permissions in the App-Specific Settings.

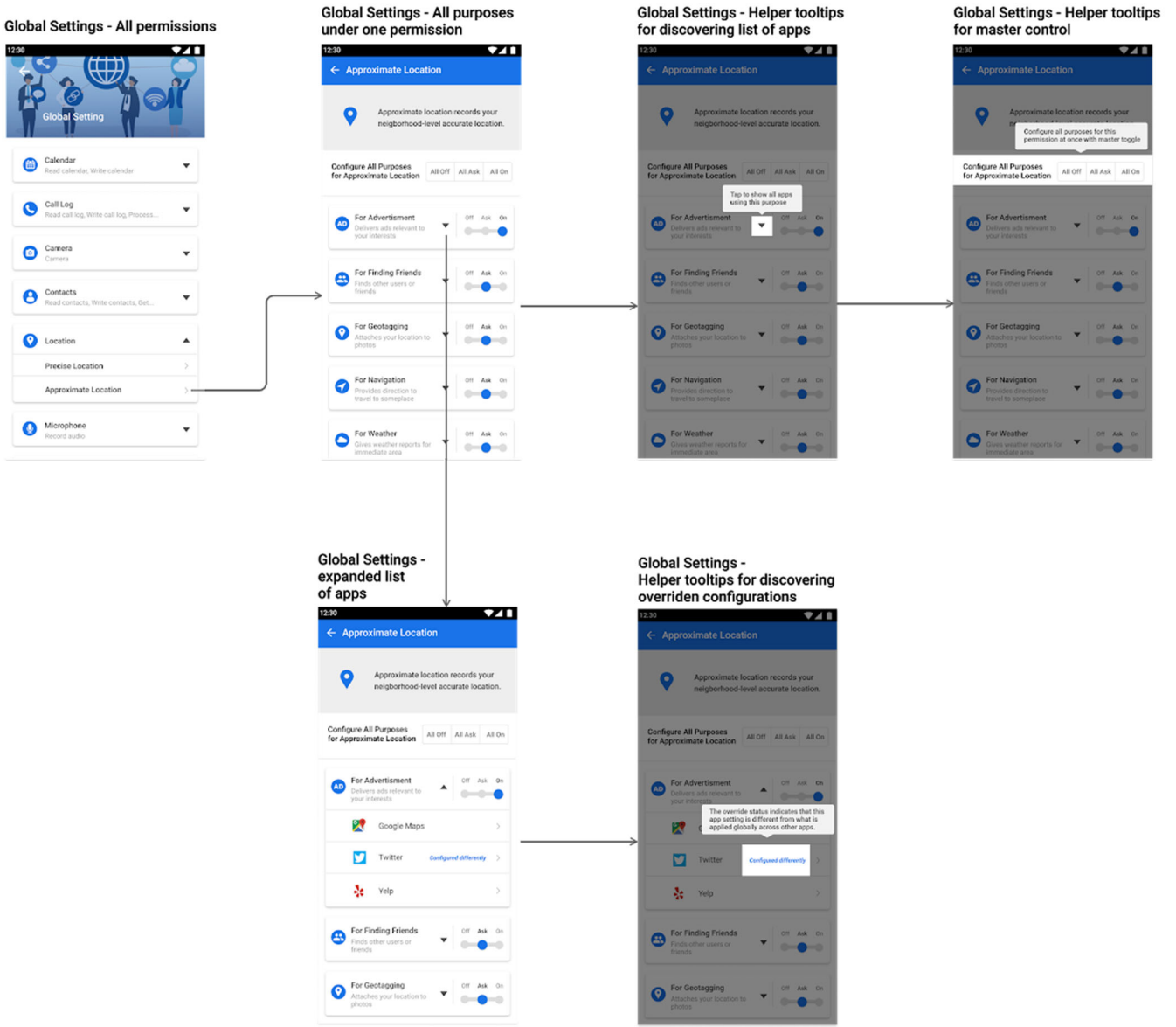

**Figure 17.** These screens show the interaction flow for Global Settings.

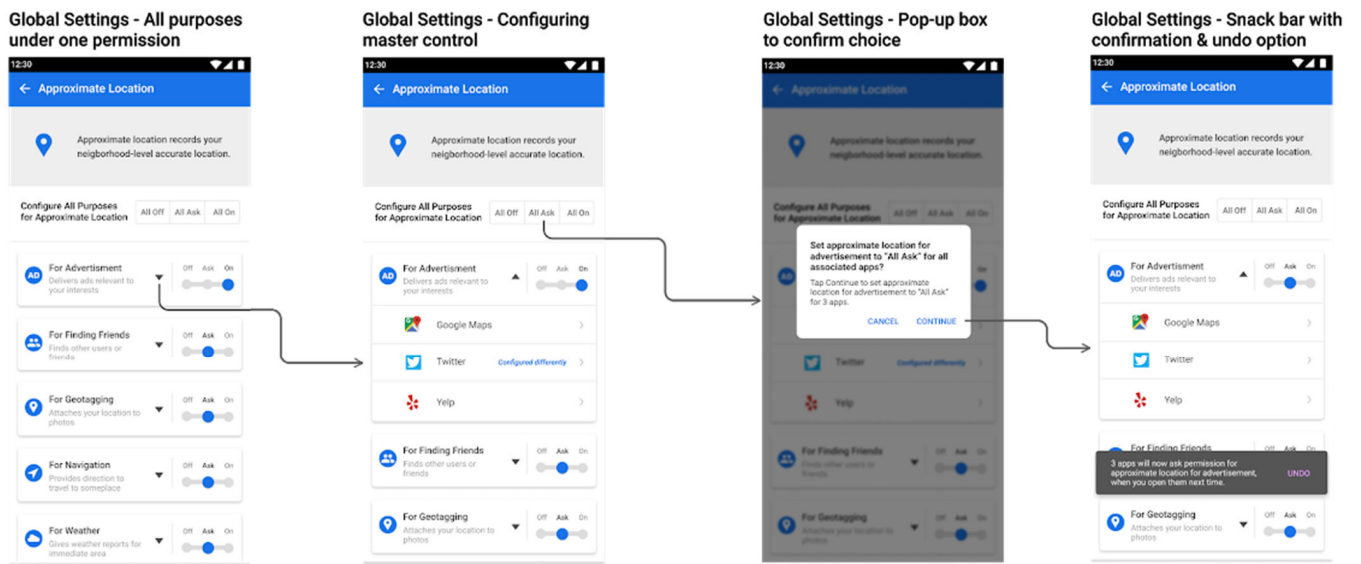

**Figure 18.** These screens show the interaction flow for configuring a single permission for all apps in Global Settings.

**APPENDIX E - SITE MAP OF USER INTERFACES FOR PE FOR ANDROID**

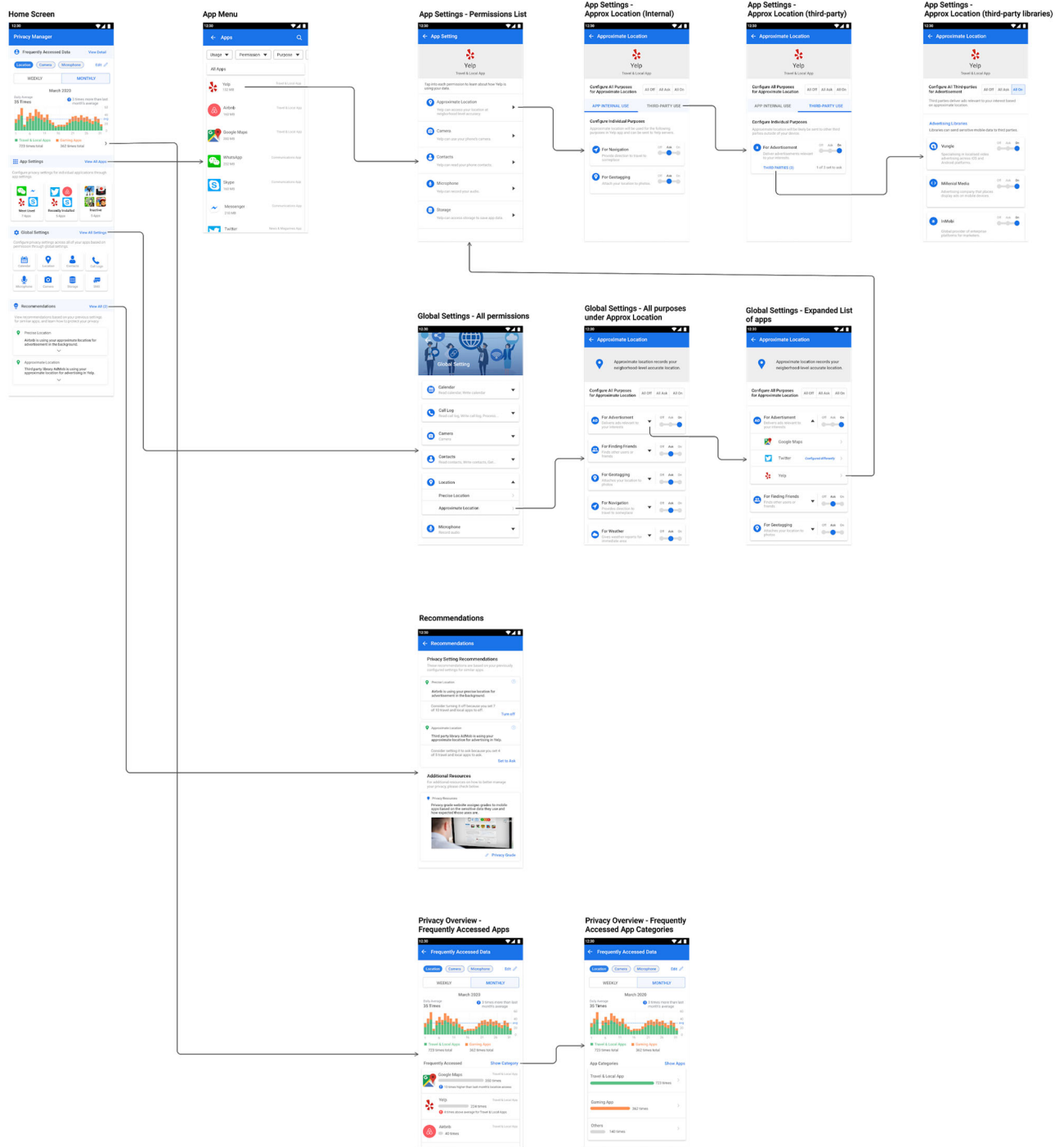

**Figure 19.** Site map starting with the Home Screen at the top left.

**GLOSSARY**

*APK* - Android Package, the package file format for mobile apps.

*App Policy* - An app policy is a json file embedded with an APK that describes the behavior of an app, in terms of permissions and purposes, and for both machines and for humans. Sometimes also referred to as an *off-device policy*. Contrast with *organization policy* and *user policy*.

*Configure time UI* - An alternative name for the *Privacy Manager*.

*Dangerous Permissions* - Dangerous permissions are the set of Android permissions which, when an app requests access to one, are checked against by the *Policy Manager* for a decision. PE for Android currently uses Android's previously publicly published list of dangerous permissions, as listed in Appendix A. Also see *Permissions* and *Sensitive Data*.

*First Party* - First party refers to the app itself (and by proxy the app developers) requesting *sensitive data*. Contrast with *Third Party*.

*Install time UI* - The install time UI is the user interface presented to end-users when installing a new app on their smartphone.

*Library* - Self-contained code created by *third parties* that can be included in an app that offers some kind of service or utility. Examples include parsing JSON or XML, game engines, or advertising. Some libraries will request sensitive data and share that data with web servers owned by the third party. Contrast with *First Party*.

*Notification UI* - The notification UI is the user interface shown in Android's notifications window shade, to inform end-users when a permission is automatically allowed or denied based on an existing *user policy* or *organization policy*.

*Off-device Policy* - An alternative name for *app policy*. Sometimes abbreviated as ODP.

*Organization Policy* - An end-user's mandatory settings for an (app, permission, purpose) tuple, in the form of allow, deny, or ask. The organization policy is set by the end-user's organization, and read in and enforced by PE for Android. Contrast with *app policy* and *user policy*.

*Permission* - Android's current mechanism for enforcing privacy, where developers are required to declare all permissions that an app will use out of a list of existing permissions. Some permissions are designated by Android as *Dangerous Permissions*. *PE for Android* extends the concept of permissions by adding *purposes* and differentiating between first-party and third-party requests for sensitive data. Also see *Dangerous Permissions*, *Purpose*, and *Sensitive Data*.

*PE for Android* - Privacy Enhancements for Android, a DARPA-sponsored project to improve end-user privacy in the Android Open Source Project.

*Policy Card* - A common way of presenting *permission* and *purposes* to end-users in the *Install Time UI*, *Runtime UI*, and *Privacy Manager*.

*Policy Manager* - The Policy Manager is the configure time user interface that end-users can use to configure their privacy settings and examine *user policies* for specific apps. The Policy Manager also allows end-users to read in an *Organization Policy*. The configure time UI is located in the smartphone's Settings screen. An alternative name sometimes used for *Privacy Manager*.

*Privacy Manager* - Also known as the *Configure TIme UI* and the *Policy Manager*.

*Purpose* - A short machine-readable value characterizing why an app is requesting sensitive data. Purpose Strings are specified in an *app policy* and presented to end-users in the *Install Time UI*, *Runtime UI*, and *Privacy Manager*, typically in a *Policy Card*. Also see *Purpose String* and *Taxonomy*.

*Purpose String* - A short human-readable description of why an app is requesting *sensitive data*. Purpose Strings are specified in an *app policy* and presented to end-users in the *Install Time UI*, *Runtime UI*, and *Privacy Manager*, typically in a *Policy Card*. Also see *Purpose*.

*Runtime UI* - A user interface that asks end-users to make a decision about whether to allow or deny a request for sensitive data when at the time an app is requesting that data. Runtime UIs are presented to end-users only if the *user policy* for that app, *permission*, and *purpose* tuple is set to "Ask". Also see *Policy Card*.

*Sensitive data* - A general term referring to data that end-users may find sensitive, for example calendar, location, and microphone. Also see *Dangerous Permissions*.

*Taxonomy* - In the specific context of this white paper, taxonomy refers to the taxonomy of *purposes* characterizing why an app is using a permission. The taxonomy of purposes is in the Appendix of this document.

*Third Party* - Third party refers to *libraries* that are included with an app that request sensitive data, often sent to the third party's servers. Examples would include advertisers. Contrast with *First Party*.

*User policy* - An end-user's discretionary settings for an (app, permission, purpose) tuple, in the form of allow, deny, or ask. Contrast with *app policy* and *organization policy*.